\documentclass[reqno,a4paper,12pt]{article}
\RequirePackage{times}
\usepackage{amsmath}
\usepackage{amssymb}
\usepackage{upref}
\usepackage[english]{babel}
\usepackage{graphicx}

\numberwithin{equation}{section}

\newcommand{\A}{\mathcal{A}}
\newcommand{\B}{\mathcal{B}}
\newcommand{\C}{\mathcal{C}}
\newcommand{\D}{\mathcal{D}}
\newcommand{\F}{\mathcal{F}}
\newcommand{\Order}{\mathcal{O}}

\title{An inventory of Lattice Boltzmann models of multiphase flows
}

\author{Erik Aurell$^{1,2,3}$
and Minh Do-Quang$^{2}$
}
\date{\today}

\begin{document}

\maketitle


{\footnotesize
\begin{tabular}{rp{11.5cm}}

        $^1$ & SICS, Swedish Institute of Computer Science AB,
        SE-164 29 Kista, Sweden \\

        $^2$ & Dept. of Numerical Analysis and Computer
        Science, Kungliga Tekniska H\"ogskolan, SE--100~44 Stockholm,
        Sweden \\

	$^3$ & NORDITA, Blegdamsvej 17, DK-2100 Copenhagen, Denmark

\end{tabular}
}

\begin{abstract}
This document reports investigations of models of
multiphase flows using Lattice Boltzmann methods. The emphasis is on deriving
by Chapman-Enskog techniques the corresponding macroscopic equations. 
The singular interface (Young-Laplace-Gauss) model is described
briefly, with a discussion of its limitations.
The diffuse interface theory is discussed in more detail,
and shown to lead to the singular interface model in the proper
asymptotic limit. 
The Lattice Boltzmann method is presented in its simplest
form appropriate for an ideal gas. Four different Lattice
Boltzmann models for non-ideal (multi-phase) isothermal flows
are then presented in detail, and the resulting macroscopic
equations derived. Partly in contradiction with the
published literature, it is found that only one of the 
models gives physically fully acceptable equations.
The form of the equation of state for a multiphase system in
the density interval above the coexistance line determines 
surface tension and interface thickness in the diffuse interface
theory. The use of this relation for optimizing a numerical
model is discussed. The extension of Lattice Boltzmann
methods to the non-isothermal
situation is discussed.
\end{abstract}



\vfill\hrule\vspace*{0.2cm}

\thanks{{\footnotesize
Work supported by the Parallel Scientific Computing Institute
(PSCI) at NADA/KTH as project 24082-61206.}
}
\newpage
\section*{Preface}
This document reports on a project carried out 
by the authors
in March-August, 2000. While waiting for renewed 
funding
the results where not properly written up, but
left in the form found here. It now being clear that
the project has definitively been discontinued, the unfinished
report is deposited as e-Print as is, in the hope 
that some of it may be of some use to someone somewhere.
\\\\ 
As a short guide, the paper contains first a survey
of the diffuse interface models, also known as
Cahn-Hilliard models.  None of this is new, but 
is here collected in one place. It is pointed out that
two different versions of the diffuse interface theory
are used in the literature. The paper thereafter contains
an introduction the Lattice Boltzmann equations along
standard lines, and then a rederivation of the
model of Swift et al. Again, nothing here is new,
but the calculations are spelled out in more detail
than in Swift et al.
\\\\ 
The paper then contains a survey of other proposed
LBE models (isothermal), chapters 5--7, where we
reach different results than in the published 
literature. In fact, we find that one such model
not only does not give the correct momentum equation,
but even displays a spurious density diffusion term
in the continuity equation! On the other hand,
we find that one other model presented in the literature
actually gives the correct continuity and momentum
equations.
\\\\
Section 8 contains a discussion of thermal multiphase
Lattive Boltzmann equations. The investigations reported
are not conclusive, and probably along the wrong path.
The general setting, and the references, could however
be of use. We believe that constructing thermal multiphase
Lattice Boltzmann equations is a difficult problem. We
doubt it can be done in the standard simplifying single
relaxation-time approximation. 
\\\\
Section 9 contains a discussion of optimization issues 
for the model described in section 7. By changing the
equation of state in the interface region, one may
change any of either three properties, keeping the others
constant. The properties are the characteristics of the
stable phases, the surface tension
and the interface thickness.
If, for instance, one wants to simulate an interface
on a given grid, then the interface thickness 
in the model must be significantly larger. Hence,
by changing only interface thickness, one could simulate
multiphase fluids with given density contrast and
surface tension on different grids. The numerical
results do not quite illustrate this fact, which
we nevertheless believe should be important in practical
use.
\\\\
The work in sections 4 and 5 were done in collaboration
with Massimo Vergassola, Observatoire de Nice, to whom
we direct our heartful thanks for patient instruction
and very useful advice on the calculations.

\tableofcontents

\section{Introduction}
\label{s:introduction}
Multiphase flows are difficult because
they involve thermodynamics (co-existing phases)
kinetics (nucleation, phase transitions), 
and hydrodynamics (inertial effects).
The nature of a liquid-vapour interface is still
partly an open problem from the physical point of view.
Mathematical descriptions of a multiphase flow
have hence the status of models, and 
can be classified as
i) singular interface models, which go
back to Young, Laplace and Gauss; ii) diffuse
interface models, which were first developed by
Maxwell, Gibbs, Van der Waals~\cite{van:thermo-78} 
and Korteweg~\cite{Korteweg}, and in a more
modern setting by Cahn~\&~Hil\-liard~\cite{cahn:interface-58},
and iii) fully detailed  statistical
mechanics models, which are still under development~\cite{Evans-79}.
\\\\
In numerical work the most common are methods
based on the first kind of models, mainly
because they do not require the resolution of 
the interface,
and can therefore be simulated with
less effort.
We begin below with a brief discussion of such models, and in which 
situations they are likely to be inadequate.
We then turn to the diffuse interface theory,
from which one
can derive the singular interface theory as
an asymptotic limit.
\\\\
If one is only interested in phenomena where
the singular interface model is expected to
be correct,
the diffuse interface theory can,
from the computational point of view,
be considered a method of solving a moving
boundary problem by solving a corresponding
set of reaction-diffusion equations.
The difference between a purely numerical procedure
and the diffuse interface theory is that the latter
is also a continuum thermodynamic
model of the interface, and can therefore be compared
with a wider set of data.
\\\\
Lattice Boltzmann methods are computational
schemes to solve macroscopic equations
by solving the equations of
a microscopic world, which is
described by the desired
macroscopic equations in
an asymptotic limit.
An review of early work in the area
is \cite{be:basic-92},
for a comparatively recent review, see
\cite{ch:lattice-98}.
If the microscopic world would be the real
world, we would have a method akin to
Molecular Dynamics, inevitably  
an expensive and cumbersome way to solve the
macroscopic equations. 
What makes Lattice Boltzmann 
an interesting method 
is that we have freedom in choice of the
microscopic model. In fact, we can
make a rather drastic departure from
the real world, and consider a microworld
where velocities only take values in
a discrete set. If such a model is
correctly set up, the resulting macroscopic equations
will still be the desired ones.
\\\\
From the PDE standpoint, Lattice Boltzmann methods
would seem to be wasteful, since they require
resolution of scales smaller than hydrodynamic.
In fact, however, Lattice Boltzmann 
methods do not require resolution
of very small scales, and should rather be
thought of as finite difference schemes with
auxiliary variables. 
In this regard
Lattice Boltzmann methods are superior to 
Lattice Gas models, which use Boolean
variables, and require an additional averaging,
and hence a significantly larger scale
separation. The fluctuations on the
other hand make Lattice Gases advantageous 
for some thermodynamic stability problems, 
such as those recently
investigated by Boghosian and coworkers 
\cite{Boghosian}.
\\\\
The competitiveness of Lattice Boltzmann
methods with standard PDE methods
-- at present, and in the future --
will mostly be
determined by how large an overhead the
extra variables in Lattice Boltzmann
represent, and by the overall
versatility and stability of the method.
We will not attempt such a comparison in this
report. A fair summary of the literature is
that the opinion is still divided, but that
Lattice Boltzmann methods have so far mainly
been used in the physics community, where 
the interpretation in terms of a fictious microworld
is considered more of
an advantage than a draw-back.
A related point of view is 
that Lattice Boltzmann methods 
are closer to the physics, in the same sense
as kinetics is compared with hydrodynamics.
In flows with complicated physics the hydrodynamical 
descriptions may not have been worked out,
or may be unfamiliar, and in both cases
using Lattice Boltzmann methods could then
be a more effective solution.
The example usually given in this context
is flow in porous media \cite{ch:lattice-98}.
\\\\
Given the reasons
why Lattice Boltzmann methods for multiphase
could be interesting, it still remains
to construct such models. In this report
we will go through models
proposed in the literature, and find them
lacking in several respects. Some of this
work, presented below in sections 4-6 
repeats calculations in the literature,
while some of it is new. In section 7
we show that one of the schemes proposed
in the literature on the other hand
gives correct 
continuity and momentum equations for an
isothermal multiphase flow. The construction
involves an auxiliary quantity, and an expansion
which mixes orders for this auxiliary quantity.§
In section 8 we show that the model of section
7 can be generalized to give non-isothermal
multiphase flows. We note that other Lattice Boltzmann
models for non-isothermal flows have been 
introduced in the literature, but do not review
them in detail. In section 9 we discuss optimizing
issues by modifying the free energy density in
the interface region.

\section{Models of multiphase flows}
\label{s:multiphase}
\subsection{The singular interface models}
In the singular interface model a surface is prescribed
separating the two phases. The free energy includes
a term proportional to the area of the surface, and
spatially located with support on the surface, hence
the name singular interface model.
\\\\
For definitiveness, assume that
the fluid obeys 
Navier-Stokes equations in both phases,
and that the interface is an impermeable
separating surface.
A necessary boundary condition is then that
the normal velocity components are continuous
across the interface
\begin{equation}
\vec v\cdot \hat n|_- = \vec v\cdot \hat n|_+ = v_n
\end{equation}
If the fluids are viscous we also have
that the tangential components are continuous:
\begin{equation}
\left(\vec v -(\vec v\cdot \hat n)\hat n\right)|_- 
=\left(\vec v -(\vec v\cdot \hat n)\hat n\right)|_+ 
\end{equation}
If the velocity gradients at the interface vanish,
so that the interface and its surroundings are locally
at rest, and if the surface tension does
not depend on the position along the surface,
then there is a pressure jump across the interface
according to Laplace's law
\begin{equation}
\label{eq:Laplace-law}
P_+ - P_- = 2\sigma {\cal K}
\end{equation}
where $\sigma$ is the surface tension
and ${\cal K}$ is the mean curvature
of the surface, oriented such that the
$+$ side is on the inside and the
$-$ side is on the outside.
If the fluid is not at rest, to the right
hand side of \eqref{eq:Laplace-law} should
be added the difference of viscous stresses,
and if $\sigma$
depends on the position on the interface,
e.g. through a dependence on temperature,
there is a further term involving the
gradient of $\sigma$
(see \cite{landau}, \S\,60).
In general, singular interface models lead to
free boundary problems with a stress balance 
at the boundary, and can include both heat
and mass flow through the interface
\cite{Delhaye}.
\\\\
Singular interface models are likely
to be incorrect if applied to processes on length
scales similar to the thickness of the interface.
The standard list of such phenomena is
(see e.g. \cite{AndersonMcFaddenWheeler}):
\begin{itemize}
\item motion near a critical point, where the
	interface thickness is large;
\item motion of a contact line between a liquid 
	and its vapour along a solid wall, see
	also \cite{Shikhmurzaev};
\item changes in topology, such as coalescence 
	of liquid droplets or
	breakup of bubbles; 
\item spontaneous formation of bubbles in
	overheated liquids (nucleation);
\item spontaneous separation into thermodynamically
	stable phases after a rapid quench from a higher
	temperature, where only one phase
	is stable (spinodal decomposition).
\end{itemize}
The first three of these examples are from near-equilibrium
statistical mechanics, while in the last two inertial
effects are weak, at least initially.
The physics of systems that are both far from equilibrium
and coupled to macroscopic transport is not systematically
known, but is likely to give rise to new phenomena.
One example could be bubble formation 
(nucleation) in the presence of strong temperature gradients,
as in a liquid close to a heated wall.

\subsection{The diffuse interface theory: statics}
In modern terms, the diffuse interface theory is
a Landau mean-field theory. In the isothermal setting,
as considered by Van der Waals and Cahn-Hilliard,
the (Helmholtz) free energy functional is assumed to
depend on density and density gradients as
\begin{equation}
{\cal F} = \int_V \rho f(\rho,T) + \frac{1}{2}K|\nabla\rho|^2 dV
\label{eq:Cahn-Hilliard-F}
\end{equation}
where $f(\rho,T)$ is the bulk free energy density (per unit
mass) of a homogeneous phase with density $\rho$ at
temperature $T$, and $K$ is a function of $\rho$ and $T$.
For simplicity we will in the following discussion
take $K$ a constant.
\\\\
A thermodynamically
stable state at fixed mass is a minimum of the grand potential
${\cal A} = {\cal F} -\mu M$, where $\mu$ is the chemical
potential per unit mass. If there is a single spatially
homogeneous global minimum at
$\bar\rho$,
the free energy is
\begin{equation}
{\cal F} = V \bar\rho f(\bar\rho,T)\quad\hbox{(homogeneous phase)}
\label{eq:Cahn-Hilliard-hom}
\end{equation}
The thermodynamic pressure is 
$P=-\frac{\partial{\cal F}}{\partial V}|_{(M,T)}$
(derivative at constant mass and temperature). The
chemical potential per unit mass is related to the free
energy by
$\mu=\frac{\partial{\cal F}}{\partial M}|_{(V,T)}$
(derivative at constant volume and temperature). 
Since $M=\bar\rho V$, (\ref{eq:Cahn-Hilliard-hom}) leads
to 
\begin{equation}
P = \bar\rho^{\,2} f'(\bar\rho) \qquad \mu = f(\bar\rho)
+ \bar\rho f'(\bar\rho)
\label{eq:Cahn-Hilliard-P-and-mu}
\end{equation}
Alternatively, if $\psi=\rho f$ is the bulk free energy per
unit volume, then 
\begin{equation}
P = \bar\rho \psi'(\bar\rho) - \psi(\bar\rho)  \qquad \mu = \psi'(\bar\rho)
\label{eq:Cahn-Hilliard-P-and-mu-psiform}
\end{equation}
If, however, the grand potential
has more than one
global minimum, i.e. if there are two
or more coexisting phases, then the mimimum of is found
by a Euler-Lagrange variational equation:
\begin{equation}
\mu = (\rho f)'_{\rho} - K \nabla^2 \rho
\label{eq:variational}
\end{equation}
In one dimension, this
equation can be integrated once to give
\begin{equation}
\label{eq:variation-integrated}
\left(\rho f - \rho\mu\right) = \frac{1}{2}K
|\partial_z\rho|^2 - P
\end{equation}
We expect there to be
solutions of these equations which consist
of domains where $\rho$ is almost constant,
and equal to one of the minima, separated
by flat interfaces. Wherever $\nabla^2 \rho$ can
be neglected, the Lagrange multiplier $\mu$
is equal to $\rho f' + f$, which by
(\ref{eq:Cahn-Hilliard-P-and-mu}) is the chemical
potential. If density varies but in one direction,
then \eqref{eq:variation-integrated} is generally
valid, and when then $|\nabla\rho|^2$ can 
be neglected, the constant of integration $P$
on the right hand side of is by 
\eqref{eq:Cahn-Hilliard-P-and-mu} 
equal to
the pressure.
The thermodynamic rules that the
chemical potential and the pressure
in two coexisting phases in equilibrium 
are equal, can then be seen as conditions
that the solutions of (\ref{eq:variational}) 
and \eqref{eq:variation-integrated} are
of the desired type.
\\\\
If we multiply (\ref{eq:variational}) with the gradient
of $\rho$, we see that it is equivalent to
\begin{equation}
\partial_{\alpha}T_{\alpha\beta} = 0 \quad
T_{\alpha\beta} = \rho(f-\mu)\delta_{\alpha\beta} - 
K\left(\partial_{\alpha}\rho\partial_{\beta}\rho -\frac{1}{2}
|\nabla\rho|^2\delta_{\alpha\beta}\right)
\label{eq:stress-tensor}
\end{equation}
where $T$ is a capillary stress tensor.
In one dimension $T$ has only one component, and
\eqref{eq:stress-tensor} reduces to \eqref{eq:variation-integrated}.

\subsection{The diffuse interface theory: dynamics}
A difuse interface model of a multiphase
flow is a set of equations 
for density ($\rho$), momentum density
($\rho u_{\alpha}$) and energy density per
unit mass ($\varepsilon$),
that reduces
to the static case just discussed, if velocity
is zero.
These equations will be of the form 
\begin{eqnarray} 
\label{eq:continuity-diffuse-interface}
\partial_t\rho + \partial_{\alpha}(\rho u_{\alpha})&=& 0 \\
\label{eq:momentum-diffuse-interface}
\rho D_tu_{\beta} &=& \partial_{\alpha}(T_{\alpha\beta}+
\sigma^{(v)}_{\alpha\beta})\\
\label{eq:energy}
\rho D_t \varepsilon &=& \partial_{\alpha}((T_{\alpha\beta}+
\sigma^{(v)}_{\alpha\beta})
u_{\beta}) -\partial_{\alpha} q_{\alpha}
\end{eqnarray}
where $T_{\alpha\beta}$ is the reversible part
of a stress tensor, $\sigma^{(v)}_{\alpha\beta}$
the hydrodynamic
viscous stress tensor, and 
$q$ an energy flux.
Using the first (continuity) equation
the left hand side of the two others can be rewritten
$\partial_t(\rho u_{\alpha}) + \partial_{\beta}(\rho u_{\alpha} 
u_{\beta})$
and $\partial_t(\rho \varepsilon) + \partial_{\alpha}(\rho\varepsilon u_{\alpha})$.
The energy density consists of two parts: a kinetic energy
($\frac{1}{2}\rho u^2$), and an internal
energy density $\rho \varepsilon_I$.
The equations (\ref{eq:continuity-diffuse-interface}), 
(\ref{eq:momentum-diffuse-interface})
(\ref{eq:energy}) must be supplemented with an equation of state,
and a constitutive relation for the energy flux,
which according to \cite{AndersonMcFaddenWheeler}
contains a heat flux, and a ``non-classical''
term of interstitial
working:
\begin{eqnarray} 
\label{eq:heat-flux}
q_{\alpha} &=& -\kappa \partial_{\alpha} T + K \rho (\nabla\cdot u)
		\partial_{\alpha}\rho
\end{eqnarray}
The appearence of the second term in the energy flux
(but not in the entropy flux) can be traced back to
an assumption by Van der Waals, that the gradient
term in the free energy in fact stems entirely from
the internal energy. It allows for some simplifications,
in that the equations for the internal energy density $\varepsilon_I$
and entropy density $s$ read
\begin{eqnarray} 
\label{eq:internal-energy-diffuse-interface}
\rho D_t \varepsilon_I &=& \partial_{\alpha}
(\kappa \partial_{\alpha} T)
-p(\nabla\cdot u) +   
\sigma^{(v)}_{\alpha\beta}\partial_{\alpha}u_{\beta} \\
\label{eq:entropy-diffuse-interface}
\rho T D_t s &=& \partial_{\alpha}
(\kappa \partial_{\alpha} T)   +
\sigma^{(v)}_{\alpha\beta}\partial_{\alpha}u_{\beta} 
\end{eqnarray}
where $p$ is a pressure, related to density and temperature
by the equation of state (\ref{eq:Cahn-Hilliard-P-and-mu}).
If $f[\rho,T,\nabla\rho]$ is the free energy functional
density in 
(\ref{eq:Cahn-Hilliard-F}) of a system at rest, 
$\varepsilon_I[\rho,T,\nabla\rho]$ the internal energy functional
density, and $s(\rho,T)$ the entropy density
, then $f=\varepsilon_I-Ts$. 
The equation of heat transport (equation
\eqref{eq:entropy-diffuse-interface})
is therefore not independent, but can be derived
from (\ref{eq:internal-energy-diffuse-interface}),
the equation of state and thermodynamic identities,
see \cite{landau} \S\,49.
The recent review
\cite{AndersonMcFaddenWheeler}
lists a number of problems in which diffuse interface
models have been used successfully, for other
recent papers, see 
\cite{na:twophase}
and \cite{JasnowVinals}. 
\\\\
Different dynamic diffuse interface models
differ by the choice of $q$ and the stress tensor $T$.
Gibbs et al \cite{Gibbs-76},
who credit Lovett and Lebowitz \& Percus,
use for $T$ the stress tensor in
(\ref{eq:stress-tensor}).
This choice has the perhaps undesired feature that
it is not entirely local, since it depends on the
chemical potential $\mu$: it assumes that some part
of the fluid is at thermal equilibrium and at rest,
so that $\mu$ can be measured there.
Another choice is to use equation 
(\ref{eq:variational}) to solve for $\mu$,
and substitute that expression
in (\ref{eq:stress-tensor}): this is the
form used in e.g. \cite{AndersonMcFaddenWheeler}.
We note that equation (\ref{eq:variational})
holds for a fluid at rest, but not necessarily for
a fluid in motion, compare (\ref{eq:momentum-diffuse-interface}).
\\\\
We will not discuss further which of the
most common forms in the literature 
gives the correct description of the dynamics
of a diffuse interface. As a historic aside, 
it is however interesting to note that 
Korteweg in his 1901 paper \cite{Korteweg}
derives by symmetry arguments
the following more general form
of the inviscid
stress tensor
(\cite{Korteweg}, p. 12, equation (20) includes
also the viscous terms)
\begin{eqnarray}
\label{eq:stress-tensor-2}
T_{\alpha,\beta} &=& 
\delta_{\alpha\beta} 
	\left(p +a |\nabla\rho|^2 - c \nabla^2\rho\right) \nonumber \\
&& +\, b\, \nabla_{\alpha}\rho \nabla_{\beta}\rho
+\, d\, \nabla^2_{\alpha\beta}\rho 
\end{eqnarray}
with four undetermined coefficient functions
($a$, $b$, $c$ and $d$) to describe the capillary forces.
Korteweg's stress tensor reduces to the one
of \cite{Gibbs-76} if one takes $p$ equal to 
$\rho(f-\mu)$, $b$ equal to $-K$, $a$ equal to $K/2$,
and $c=d=0$. Korteweg further derives a
surface tension (compare (\ref{eq:surface-tension1})
\begin{equation}
\label{eq:surface-tension2}
\sigma = \int \left(b + \frac{\partial d}{\partial\rho}\right)
| \frac{\partial\rho }{\partial z} |^2 dz
\end{equation}
which shows that the diagonal components of the
stress tensor do not show up in the surface tension.
\\\\
We end this brief introduction by discussing the
limits of validity of the  Cahn-Hilliard
free energy (\ref{eq:Cahn-Hilliard-F}). If $K$ is
a constant, it follows that the surface tension
decreases to zero  
along the critical line as 
$\sigma\sim\left(T_c-T\right)^{\mu}$
with $\mu=\frac{3}{2}$.
The correct
value of this critical exponent
is however $\mu\approx 1.28$~\cite{Widom-72},
which reflects the fact that a Landau 
theory does not correctly
describe this second order phase transition.
To match the observed dependence of the thickness
one may take $K$ a parametric function of temperature
and density. It is still however an open question
if such a modified Cahn-Hilliard theory correctly
describes the interface close to a critical point.
\\\\
Along the coexistence line away from the critical
point, the Cahn-Hilliard theory predicts a sharp
interface, on the order of a few atomic diameters.
This is in agreement with experiments,
but nevertheless a conceptual problem, because if the
interface is so sharp there is no reason why higher
order terms should be relatively small.
Since the singular interface theory can be derived
from the diffuse interface theory, and since both
have been applied to widely different flows, 
there should however be 
some meaning to truncating (\ref{eq:Cahn-Hilliard-F})
after the first gradient term, perhaps as some sort
of normal form.
For a detailed discussion of gradient expansion
resummation techniques of first principle
statistical mechanics models, and of possible
problems with the diffuse interface models which
go outside the range of phenomena of interest to us here, 
see \cite{Evans-79}.

\subsection{Derivation of the singular interface theory}
{\small
For completeness we derive here in the simplest case
a singular interface model from a diffuse interface theory.
\\\\
If $K$ is small and density gradients are close to
zero everywhere but in a small part of the volume, then
that part can be considered a fuzzy interface. 
Let us suppose that this interface passes through point $P$
and is there approximated by a surface given parametrically
by $z = \frac{1}{2}(r_1x^2+ r_2y^2)$. The two principal
components of the curvature tensor ($r_1$ and $r_2$) need
not have the same sign. 
Let us further assume that density is only a function
of $z - \frac{1}{2}(r_1x^2+ r_2y^2)$.
A unit vector
normal to the interface is formed by the density gradient 
\begin{equation}
\hat n = {{\nabla\rho}\over{|\nabla\rho|}} 
= {{(-r_1 x,-r_2 y,1)}\over
   {\sqrt{1+r_1^2 x^2+r_2^2 y^2}}}
\end{equation}
and points for small $x$ and $y$  
approximately in the $\hat z$ direction.
The stess tensor reads 
\begin{equation}
\sigma_{\alpha\beta} = \left(-P+\frac{1}{2}p_1\right)
\delta_{\alpha\beta} - p_1\hat n_{\alpha} \hat n_{\beta}
\end{equation}
where $P = \rho(\mu-f)$ is the thermodynamic 
pressure in a homogeneous phase at density $\rho$,
and $p_1$ is $K(\frac{\partial\rho}{\partial a})^2$,
evaluated at $a = z - \frac{1}{2}(r_1x^2+ r_2y^2)$.
\\\\ 
On the $z$ axis the only non-zero component of
the force field $\partial_{\alpha}\sigma_{\alpha\beta}$ is in the
$z$ direction:
\begin{equation}
\partial_{\alpha}\sigma_{\alpha\beta}(0,0,z) = \hat z
\left(-\partial_z P -\partial_z p_1+ p_1(r_1+r_2)\right) 
\end{equation}
On the other hand, we know that at rest $\partial_{\alpha}
\sigma_{\alpha\beta}=0$.
If $z_0$ and $z_Q$ are two points on the $z$ axis
on opposite sides of the interface we therefore
have
\begin{equation}
0=\int_{z_O}^{z_Q} \partial_{\alpha}\sigma_{\alpha\beta}(0,0,z)
dx_{\beta} = 
-(P(z_Q)-P(z_O)) + (r_1+r_2) \int_{z_O}^{z_Q} K|\partial_z\rho|^2 dz
\label{eq:surface-tension1}
\end{equation}
The pressure difference is proportional
to the mean curvature ($\frac{r_1+r_2}{2}$) and to the
gradient contribution to the free energy in 
(\ref{eq:Cahn-Hilliard-F}). We can therefore identify
that term as a free energy proportional to the area 
of the surface, i.e. a surface tension.

}

\section{Lattice Boltzmann equation models in brief}
\label{s:LBE-in-brief}
A physical system can be described on different
levels, each of which corresponding to different resolutions
in time and space. In statistical mechanics of 
essentially classical systems, such as ordinary liquids
and gases, there are three important levels of
description. The first (most detailed) level considers
the motion of individual atoms and molecules. In a real
liquid at room temperature and atmospheric
pressure this implies spatial scale
of about $10^{-9}$ m, and temporal scale of about 
$10^{-12}$ s. Considerations at this scale are important
e.g. in modelling chemical reactions in solution.
\\\\
The second level of description becomes valid when the actual
motion can be substituted with the mean number of particles
to have momentum in a range $[\vec \xi,\vec \xi+\Delta \vec \xi]$ at positions
$[\vec x,\vec x+\Delta \vec x]$. The condition for this level
to be accurate is that fluctuations of this number are relatively
small, or that the mean number is much larger than one.
The system is then described by a chain of distribution
functions, where $f(\vec \xi, \vec x, t)$ is the probability density
of finding a particle at position $\vec x$ with momentum $\vec \xi$
at time $t$, $f_2(\vec \xi_1,\vec \xi_2, \vec x_1,\vec x_2, t)$
is the joint probability of finding one particle at $\vec x_1$
with momentum $\xi_1$ and another at $\vec x_2$ with momentum $\xi_2$,
and so on.
A description of a gas or a liquid in terms of distribution
functions is called kinetic theory. The simplest such
description, valid for a very rare gas, is Boltzmann's
equation, which in the single relaxation time approximation 
reads:
\begin{equation} 
\label{eq:basic-boltzmann}
  \frac{\partial f}{\partial t}+\xi.\vec{\nabla}f=-\frac{1}{\tau}(f-f^{eq})
\end{equation}
where $f^{eq}$ would be the distribution function of a liquid or
a gas in termal equilibrium with the same density, velocity and
temperature.
Boltzmann's equation can include a force field (acceleration term)
and of course Boltzmann's full
collision operator $\Omega$, quadratic in $f$.
In this paper we will however in accordance with most
of the Lattice Boltzmann equation literature stay with the
(simpler) single relaxation time approximation for $\Omega$.
\\\\
The first three
moments of the distribution function $f$ are
respectively the density ($\rho$), the momentum density,
($\rho u$)
and the kinetic energy density per unit mass ($\varepsilon_K$):
\begin{subequations}
\begin{align}
  \int f(\vec \xi, \vec x, t) d\, \vec \xi &= \rho(\vec x, t) \\
  \int \vec \xi f(\vec \xi, \vec x, t) d\, \vec \xi &= 
	\rho(\vec x, t)\,u(\vec x, t)\\
  \int \frac{\xi^2}{2m} f(\vec \xi, \vec x, t) d\, \vec \xi &= 
	\rho(\vec x, t)\, \varepsilon_K(\vec x , t)
\end{align}
\end{subequations}
where $m$ is the mass of a particle.
We can introduce a local
temperature field $T$ such
that $\varepsilon_K= \frac{1}{2}u^2+\frac{D}{2} R T$,
where $D$ is dimension of space (2 or 3 
in the cases of interest to us)
and $R$ the gas constant.
The space integrals of density, momentum density
and total energy density are integrals of the motion,
namely total mass, total momentum and total energy.
Large-scale modulations of these densities
in space therefore have slow dynamics in time, 
compared to the other degrees
of freedom on the kinetic level.
A hydrodynamics of a physical system refers in general
to a description only in terms of such slow,
large-scale modes \cite{ChaikinLubensky}, which
are therefore called hydrodynamic modes.
Hydrodynamics in the ordinary sense, e.g.
to describe the motion of water, is an example
of such a description, albeit more complicated than
for some other systems, 
since total energy is implicitly
given by velocity, density and temperature through
the equation of state.
\\\\
A Lattice Boltzmann equation is a kinetic level 
description of a fictitious microworld with the same
conserved quantities as the real world. If properly
constructed (see below), it will then
give rise to the same hydrodynamics. The smallest 
scale that needs to be resolved to compute the hydrodynamic
equations in this manner is only such that hydrodynamics
is valid for the scale of the object of interest.
For practical purposes, Lattice Boltzmann equations 
are therefore a sort of finite difference schemes,
which contain
auxiliary variables that stand for purely kinetic (not hydrodynamic)
modes. The number of auxiliary modes is typically 2-5
more than the number of hydrodynamic modes, which
represents an overhead.
\\\\
A quick way to introduce the Lattice Boltzmann equations
is to discretize space on a regular lattice, and momentum to
a set of vectors in the dual lattice. Equation 
(\ref{eq:basic-boltzmann}) can then be solved to first-order
accuracy in time
\begin{equation} \label{eq:boltzmann-discrete}
  f_i(\vec x+e_i,t+\Delta t)-f_i(x,t) = -\frac{1}{\tau}(f_i-f_i^{eq})
\end{equation}
where the indices $i$ label the 
set of momentum vectors, $f_i$ is 
the probability to be at lattice point $\vec x$
with momentum $e_i$ at time $t$,
and time $\Delta t$ is such
that a particle with velocity $e_i$ moves from $\vec x$
to $\vec x + e_i$.
The hydrodynamic variables are defined in a way similar to
the continuum, i.e. 
\begin{subequations}
\label{eq:macro}
\begin{align}
\label{eq:macro-conservation}
  \sum_i w_i f_i &= \rho \\
\label{eq:macro-momentum}
  \sum_i w_i f_i e_i &= \rho u \\
\label{eq:macro-energy}
 \frac{1}{2}  \sum_i w_i f_i |e_{i}-u|^2 &= \rho \frac{D}{2} RT
\end{align}
\end{subequations}
where the $w_i$ are a set of weights. 
A given Lattice Boltzmann model is then determined
by equation (\ref{eq:boltzmann-discrete}) (or a generalization
thereof), and how the equilibrium distribution function
$f_i^{eq}$ depends on $\rho$, $u$ and $T$.
\\\\
The a priori conditions for the same hydrodynamical
equations
to result from a Lattice Boltzmann model as from 
a real physical system, is that the symmetry of
the lattice is sufficiently close to the full rotational
symmetry of space.
To capture correctly dissipative terms one needs that
the invariant tensors form from the lattice vectors
up to fourth order are isotropic:
\begin{subequations}
\begin{align}
  \sum_i w_i  &= 1 \\
  \sum_i w_i e_{i\alpha} &= 0 \\
	\label{eq:invariant-second-order}
  \sum_i w_i e_{i\alpha} e_{i\beta} &= \Upsilon^{(2)}\delta_{\alpha\beta} \\
  \sum_i w_i e_{i\alpha} e_{i\beta} e_{i\gamma} &= 0 \\
	\label{eq:invariant-fourth-order}
  \sum_i w_i e_{i\alpha} e_{i\beta} e_{i\gamma} e_{i\theta} &=
            \Upsilon^{(4)}(\delta_{\alpha\beta}\delta_{\gamma\theta} +
                         \delta_{\alpha\gamma}\delta_{\beta\theta} +
                         \delta_{\alpha\theta}\delta_{\beta\gamma}) 
\end{align}
\end{subequations}
Greek indices ($\alpha, \beta ...$) here label spatial directions,
$i$ the lattice vectors and $\Upsilon^{(2)}$ and $\Upsilon^{(4)}$
are lattice constants.
Natural choices of lattices share with space the property that
all odd order invariant tensors are zero, and the condition
on second-order is easily satisfied.
The technical difficulty overcome in the
80'ies (in the context of Lattice Gases) was to choose
lattices with isotropic fourth order tensors,
such as the hexagonal lattice in two dimensions. 
For faster convergence
to the hydrodynamics it has sometimes been proposed to use
lattices with isotropic sixth or higher order tensors
\cite{qian-orszag,chen-ohashi-akiyama}.
\\\\
One presently popular
model uses a lattice of $9$ velocities
in $2$ dimensions, and is therefore known as the $D2Q9$
lattice. The velocities are the rest state (velocity $e_0=0$),
along the vectors to the  
four nearest neighbours in a simple cubic lattice
(velocities $e_1$--$e_4$), and along the vectors
to the four next-nearest neighbour ($e_5$--$e_8$).
If lattice length is $c \Delta t$, the velocities and weights are
given in the following table:
\begin{center}
\begin{tabular}{|l|l|l|} \hline
Direction & Lattice vector & Weight \\  \hline
  $e_0$   & $(0,0)$   & $w_0 = 4/9$  \\
  $e_1$   & $c(1,0)$  & $w_1 = 1/9$  \\
  $e_2$   & $c(0,1)$  & $w_2 = 1/9$  \\
  $e_3$   & $c(-1,0)$ & $w_3 = 1/9$ \\
  $e_4$   & $c(0,-1)$ & $w_4 = 1/9$  \\
  $e_5$   & $c(1,1)$  & $w_5 = 1/36$ \\
  $e_6$   & $c(-1,1)$ & $w_6 = 1/36$ \\
  $e_7$   & $c(-1,-1)$& $w_7 = 1/36$ \\
  $e_8$   & $c(1,-1)$ & $w_8 = 1/36$ \\ \hline  
\end{tabular}
\end{center}
It is easily checked that all odd order invariant tensors
formed from this lattice indeed vanish, and that
is satisfies (\ref{eq:invariant-second-order})
and (\ref{eq:invariant-fourth-order})
with $\Upsilon^{(2)} = \frac{c^2}{3}$
and $\Upsilon^{(4)} = \frac{c^4}{9}$.
\\\\
For the simplest applications it is sufficient
to choose the equilibrium distribution function
as
\begin{equation}
\label{eq:equilibrium-distribution-function-standard}
  f_i^{eq} = \rho\left[1+\frac{(e_i.{\bf u})} {\Upsilon^{(2)}}+\frac{(e_i.{\bf u})^2}{2\Upsilon^{(4)}} -\frac{{\bf u}^2} {2\Upsilon^{(2)}}\right]
\end{equation}
from which we can derive
\begin{subequations}
\begin{align}
	\label{eq:zero-moment}
  \sum_i w_i f_i^{eq} &= \rho \\
	\label{eq:first-moment}
  \sum_i w_i f_i^{eq} e_{i\alpha}&= \rho u_{\alpha} \\
	\label{eq:second-moment}
  \sum_i w_i f_i^{eq} 
	e_{i\alpha} e_{i\beta} &= 
	\Upsilon^{(2)}\rho \delta_{\alpha\beta}
	+ \rho u_{\alpha}u_{\beta} \\
	\label{eq:third-moment}
  \sum_i w_i f_i^{eq} e_{i\alpha} e_{i\beta} e_{i\gamma} &= 
            \frac{\Upsilon^{(4)}}{\Upsilon^{(2)}}\rho
(\delta_{\alpha\beta}u_{\gamma} +
                         \delta_{\alpha\gamma}u_{\beta} +
                         \delta_{\beta\gamma}u_{\alpha})
\end{align}
\end{subequations}
From (\ref{eq:second-moment}) it is natural
to identify $\Upsilon^{(2)}\rho$ with a pressure
term $p$. This means that the velocity of sound,
$v_s=\sqrt{\frac{\partial p}{\partial \rho}}$,
should be $\sqrt{\Upsilon^{(2)}}$ (a constant).
For the $D2Q9$ models this constant is $c/\sqrt{3}$. 
The temperature $R T$ is on the other hand
$\frac{1}{D\rho}\sum_i w_i f_i^{eq} 
	(e_{i\alpha}-u_{\alpha})
(e_{i\alpha}-u_{\alpha})$,
a constant equal to $\Upsilon^{(2)}$.
A model using
\eqref{eq:basic-boltzmann} and
(\ref{eq:equilibrium-distribution-function-standard})
can therefore only describe isothermal hydrodynamics
with an ideal gas equation of state, 
$p= R T \rho$.
This is sufficient for deriving e.g. incompressible
Navier-Stokes dynamics in a small Mach number
expansion, but otherwise somewhat limited.
Two important issues in the following will be to
generalize 
(\ref{eq:boltzmann-discrete}) or
(\ref{eq:equilibrium-distribution-function-standard})
or both, to allow for flows which do not have
the ideal gas equation of state, and which are
not restricted to be isothermal.
\\\\
The usual presentation of Lattice Boltzmann
methods (as above) uses a uniform grid in space.
This limitation is not a principal one, by 
varying in an
appropriate manner the weights $w_i$ it is possible
to use also non-uniform grids. These issues have
not been investigated extensively in the literature,
see however \cite{na:thelattice-92} 
and \cite{xi:some-96}.

\section{Non-equilibrium thermodynamics model}
This model was
developed by Swift and collaborators
\cite{sw:lattice-95,sw:lattice-96}. 
The model has later been used in numerical
work by several teams, for contributions from
authors of this report, see \cite{no:numerical-99}
and \cite{mi:thesis-99}.
The basic idea is to assume that the distributions relax
to a local equilibrium distribution function, which in a
one-phase region coincides
with the equilibrium in ordinary one-phase flows. In the interface region the 
local equilibrium is however assumed 
to have a pressure (stress) tensor as in the
diffuse interface theory. 
The model uses the D2Q9 lattice, equation 
\eqref{eq:boltzmann-discrete} and assumes
an isothermal situation.
The equilibrium distribution function is
a generalization of (\ref{eq:equilibrium-distribution-function-standard})
\begin{equation}
\label{eq:basic_equilibrium}
\left\{
\begin{array}{ll}
  f_i^{eq} = A+Be_{i\alpha}u_\alpha+Cu^2+Du_\alpha u_\beta e_{i\alpha} e_{i\beta}+
             G_{\alpha\beta}e_{i\alpha}e_{i\beta}\\
  f_0^{eq} = A_0 + C_0u^2
\end{array}
\right.
\end{equation}
The coefficients $A$, $B$, $C$, $D$, $G_{\alpha,\beta}$
and $A_0$ and $C_0$ are chosen such that the
zeroth and first
weighted moments are $\rho$ and $\rho u$, while
the second moment is prescribed to be
\begin{equation} 
\label{eq:presstensor}
  \sum_i w_i f_i^{eq} 
	e_{i\alpha}e_{i\beta} = 
	P_{\alpha\beta} + \rho u_\alpha u_\beta
\end{equation}
where second order tensor $P$ (in 
\cite{sw:lattice-95} referred to as a non-equilibrium pressure
tensor) is chosen to be a stress tensor in a diffuse
interface model:
\begin{equation} \label{eqn:pressuretensor0}
  P_{\alpha\beta}(\vec{r}) = p(\vec{r})\delta_{\alpha\beta}+
       \kappa\partial_\alpha\rho\partial_\beta\rho 
\end{equation}
The diagonal component 
$p(\vec{r})$ of the tensor $P$ is 
\begin{equation}
\label{eqn:pr}
p(\vec{r}) = p_0-\kappa\rho\nabla^2\rho-\frac{\kappa}{2}|\vec{\nabla}\rho|^2
\end{equation}
and $p_0$ is the pressure in
a Van der Waals equation of state
\begin{equation}
  p_0  = \frac{\rho RT}{1-b\rho}-a\rho^2
\end{equation}
where $T$ is temperature (assumed constant).
The partial derivatives of the density are evaluated
as discrete differences over neighbouring lattice points.
In terms of a real system, this introduces a coupling
between neighbouring sites, i.e. a potential energy.
\\\\
The kinetic energy density computed from
the distribution function \eqref{eq:basic_equilibrium}
would in fact by
(\ref{eq:presstensor}) give rise
to a local temperature field $T(\vec r) 
= \frac{p_0}{\rho} - \kappa \nabla^2\rho$.
The equilibrium distribution in this model
is however computed not with this variable
temperature, but with 
a global constant parameter. The use
of the model is therefore restricted to
phenomena where the back action of a temperature
gradient on density and momentum can be neglected,
basically a isothermal setting.
\\\\
The model starting from (\ref{eqn:pressuretensor0})
will give rise to Gallilean non-invarian terms in
the hydrodynamic equations.
It will become clear in the following that of the various
discretization errors that appear in the Swift et al. model, 
there are three which are
more serious, and
should be considered of the same order
as the diffusive/viscous terms.
To the model will therefore be added a
term $\F_{\alpha\beta}$ 
which eliminates two of these.
\begin{equation} \label{eqn:pressuretensor}
  P_{\alpha\beta}(\vec{r}) = p(\vec{r})\delta_{\alpha\beta}+
       \kappa\partial_\alpha\rho\partial_\beta\rho +\F_{\alpha\beta}
\end{equation}
\vspace{0.5cm}\noindent
One result of of \cite{sw:lattice-96} is that the
mass and momentum equations are:
\begin{equation}
\label{eq:NS-mass}
  \frac{\partial\rho}{\partial t} + \nabla.(\rho{\bf u}) = 0
\end{equation}
and
\begin{equation}
\label{eq:NS-momentum}
\begin{split}
  \frac{\partial(\rho {\bf u})}{\partial t} + ({\bf u}.\nabla)(\rho{\bf u}) = -\nabla p_0
    + \nu\Delta(\rho{\bf u})
    +\nabla.[\lambda(\rho)\nabla.(\rho{\bf u})] \\
    -\delta_t\frac{dp_0}{d\rho}\nabla.(\nabla.(\rho{\bf u})+(\rho{\bf u}).\nabla)
\end{split}
\end{equation}
in which the kinematic viscosity $\nu$ and bulk viscosity $\lambda$ of the fluid are
determined by:
\begin{align*}
  \nu =& \delta_t \frac{2\tau-1}{2}
	\frac{\Upsilon^{(4)}}{\Upsilon^{(2)}} \\
  \lambda(\rho) =& \delta_t\left (\tau-\frac 1 2\right )\left 
	(\frac{2\Upsilon^{(4)}}{\Upsilon^{(2)}}-
	\frac{\partial p_0}{\partial\rho}\right )
\end{align*}
In fact, as we will see, equation \eqref{eq:NS-momentum} 
is only correct with the
additional assumption that the 
fluid is close to incompressible,
something which is also found in \cite{sw:lattice-95}.

\subsection{Method of successive approximations}
Following \cite{sw:lattice-95} we will derive
continuity equation and the Navier-Stokes equation
by the method of successive approximations.
The discussion in this section does not
contain other material than in 
\cite{sw:lattice-95}, but is in places carried out in
more detail.
\\\\
We start 
from equation (\ref{eq:boltzmann-discrete}) with a Taylor
expansion of the l.h.s:
\begin{equation}\label{eq:approx0}
  \Omega_i \equiv -\frac{1}{\tau}(f_i-f_i^{eq})=\sum_{k=1}^{\infty}\frac 1 {k!}\delta_t^k
       (\partial_t+e_{i\alpha}\partial_\alpha)^kf_i
     = \sum_{k=1}^{\infty}\frac 1 {k!}\delta_t^k \D^kf_i
\end{equation}
with $\D\equiv(\partial_t+e_{i\alpha}\partial_\alpha)$. 
We assume that the distribution function $f$ is close
to the equilibrium distribution function $f^{eq}$
and expand in $\delta_t$. 
\begin{equation}\label{eq:approx1}
  f_i = f_i^{eq} + (f_i - f_i^{eq}) = f_i^{eq} - 
        \tau\left(\delta_t\D f_i+\frac{\delta_t^2}{2}\D^2f_i\right) + \Order(\delta_t^3)
\end{equation}
Substituting \eqref{eq:approx1} to r.h.s of \eqref{eq:approx0} and retaining terms to 
order $\delta_t^2$
\begin{align}\label{eq:approx2}
   \frac{\Omega_i}{\delta_t} =& \D\left[f_i^{eq}-\tau\left(\delta_t\D f_i
                              +\frac{\delta_t^2}{2}\D^2f_i\right)\right]\nonumber\\
           &+\frac{\delta_t}{2}\D^2\left[f_i^{eq}-\tau\left(\delta_t\D f_i
           +\frac{\delta_t^2}{2}\D^2f_i\right)\right]+\Order(\delta_t^2)
           \nonumber \\
         =&\D f_i^{eq} - \tau\delta_t\D^2 f_i 
           + \frac{\delta_t}{2}\D^2 f_i^{eq}+\Order(\delta_t^2) \nonumber \\
         =&\D f_i^{eq} - \delta_t\left(\tau-\frac 1 2\right )\D^2 f_i^{eq}+\Order(\delta_t^2)
\end{align}
So
\begin{equation} \label{eq:approximation}
  \frac{\Omega_i}{\delta_t} = (\partial_t+e_{i\alpha}\partial_\alpha)f_i^{eq}
        -\delta_t\left (\tau-\frac 1 2\right )(\partial_t
        +e_{i\alpha}\partial_\alpha)^2f_i^{eq}+\Order(\delta_t^2)
\end{equation}
By projecting on the hydrodynamic modes we will now derive the macroscopic equations to lowest
and next lowest order in $\delta_t$.
To obtain the continuity equation \eqref{eq:NS-mass} to lowest order, we sum over $i$ and
use \eqref{eq:macro-conservation} and \eqref{eq:macro-momentum}, which gives
\begin{subequations}
\begin{equation}
  0 =  \sum_iw_i\frac{\Omega_i}{\delta_t} = 
       \partial_t\sum_i w_if_i^{eq}+\partial_\alpha\sum_i w_i e_{i\alpha}f_i^{eq}-\Order(\delta_t)
\end{equation}
Or
\begin{equation} \label{eq:continuity1}
  \partial_t\rho+\partial_\alpha(\rho u_\alpha) = \Order(\delta_t)
\end{equation}
\end{subequations}
To the next order in $\delta_t$:
\begin{equation}
  =\delta_t\left(\tau-\frac 1 2\right)\left[\partial_t\partial_t\sum_iw_i f_i^{eq}+
                2\partial_t\partial_\alpha\sum_i w_i e_{i\alpha}f_i^{eq}+
                \partial_\alpha\partial_\beta\sum_i w_i e_{i\alpha}e_{i\beta}f_i^{eq}\right]
\end{equation}
\begin{equation} \label{eq:odt1}
  = \delta_t\left(\tau-\frac 1 2\right)\left[
                \partial_t\left(\partial_t\rho+\partial_\alpha(\rho u_\alpha)\right)
                + \partial_\alpha\left(\partial_t(\rho u_\alpha)+\partial_\beta\sum_iw_i e_{i\alpha}e_{i\beta}f_i^{eq}\right)\right]
\end{equation}
Substituting in eqn.(\ref{eq:odt1}), eqn.(\ref{eq:continuity1}) and (\ref{eq:momentum0}) 
derived below we see that the $\Order(\delta_t)$ 
terms are actually $\Order(\delta_t^2)$. 
The continuity equation is therefore:
\begin{equation} \label{eq:continuity}
  \partial_t\rho+\partial_\alpha(\rho u_\alpha) = \Order(\delta_t^2)
\end{equation}
This means that there will not be any density diffusion terms in our equations, as
indeed there should not be on general grounds \cite{landau}.
\\\\
To obtain the momentum equation, (\ref{eq:NS-momentum}), we multiply
eqn.(\ref{eq:approximation}) by $e_{i\beta}$ and sum over $i$:
\begin{align}
  0 =& \partial_t\sum_i w_i e_{i\beta}f_i^{eq}+
        \partial_\alpha\sum_i w_i
        e_{i\alpha}e_{i\beta}f_i^{eq}-\Order(\delta_t)\\
\label{eq:momentum0}
  0 =& \partial_t(\rho u_\beta)+\partial_\alpha(P_{\alpha\beta} +
               \rho u_\alpha u_\beta)-
               \delta_t\left(\tau-\frac 1 2\right) \\
    &   \left[\underbrace{\partial_t^2(\rho u_\beta)}_{\bf \A_\beta}+
            \underbrace{\partial_\alpha\partial_\gamma
            \left(\sum_i w_i e_{i\alpha}e_{i\beta}e_{i\gamma}f_i^{eq} \right)}_{\bf \B_\beta}+
            \underbrace{2\partial_t\partial_\alpha(P_{\alpha\beta} +
            \rho u_\alpha
            u_\beta)}_{\bf\C_\beta}\right]+\Order(\delta_t^2)\nonumber
\end{align}
To lowest order we get the Euler equation which can be substituted back into the last parenthesis
in eqn.\eqref{eq:odt1}, to prove that the continuity equation holds to order $\delta_t^2$.
Expanding $P_{\alpha\beta}$ by 
equation \eqref{eqn:pressuretensor} we have
\begin{equation}  \label{eq:momentum1}
\begin{split}
  \partial_t(\rho u_\beta)+\partial_\alpha(\rho u_\alpha u_\beta) =
      -\partial_\beta p_0+\kappa\partial_\beta(\rho\Delta\rho)
      +\frac\kappa 2\partial_\beta|\nabla\rho|^2 \\
      -\partial_\alpha(\kappa\partial_\alpha\rho\partial_\beta\rho)
      -\partial_\alpha\F_{\alpha\beta}
      +{\bf\A_\beta}+{\bf\B_\beta}+{\bf\C_\beta}
\end{split}
\end{equation}
Equation \eqref{eq:momentum1} shows terms on the Euler level,
where some terms have been simplified using \eqref{eq:presstensor}. We will now simplify
further the three viscous-level terms denoted ${\bf\A_\beta},{\bf\B_\beta}$ and ${\bf\C_\beta}$.
Term ${\bf\A_\beta}$ can be computed by substituting back eqn.(\ref{eq:momentum1})
\begin{align}
  \partial_t^2(\rho u_\beta) &= \partial_t(\partial_t(\rho u_\beta)) \nonumber \\
    &=-\partial_t(\partial_\alpha(P_{\alpha\beta} + \rho u_\alpha u_\beta))
      +\Order(\delta_t)
\end{align}
In the following manipulations, we always imply all equalities to hold up to terms of order $\delta_t$
. Therefore,
\begin{align}
 {\bf\A_\beta}+{\bf\C_\beta} =& \partial_t(\partial_\alpha(P_{\alpha\beta}
     +\rho u_\alpha u_\beta))\nonumber \\
  =&  \partial_t\partial_\beta p_0
     -\partial_t\partial_\beta(\kappa\rho\Delta\rho)
     -\partial_t\partial_\beta\left(\frac\kappa 2|\nabla\rho|^2\right)+\nonumber\\
   &  \partial_t\partial_\alpha\left(\kappa\partial_\beta\rho\partial_\alpha\rho\right)
     +\partial_\alpha\partial_t(\rho u_\alpha u_\beta)+\partial_\alpha\partial_t\F_{\alpha\beta}
\end{align}
We see that there are very many terms to order $\delta_t$. It will be useful to concentrate on
the terms with smallest number of derivatives acting on the density. We will therefore, following 
Swift, neglect terms with as least as many derivatives as these:
\begin{align*}
  \partial_t\partial_\beta(\kappa\rho\Delta\rho) &\approx 0 \\
  \partial_t\partial_\beta\left(\frac\kappa 2|\nabla\rho|^2\right) &\approx 0 \\
  \partial_t\partial_\alpha\left(\kappa\partial_\beta\rho\partial_\alpha\rho\right) &\approx 0 \\
  \partial_\alpha\partial_t\F_{\alpha\beta} &\approx 0
\end{align*}
With this simplification,
\begin{equation}\label{eq:A+C}
  {\bf\A_\beta}+{\bf\C_\beta} =\underbrace{\partial_t\partial_\beta p_0}_{\bf AC1_\beta}
                    +\underbrace{\partial_\alpha[u_\beta\partial_t(\rho u_\alpha)]}_{\bf AC2_\beta}
                    +\underbrace{\partial_\alpha[(\rho u_\alpha)\partial_t u_\beta]}_{\bf AC3_\beta}
\end{equation}
These terms can be further simplified by
\begin{equation} \label{eq:AC1a}
  {\bf AC1_\beta} = \partial_t\partial_\beta p_0 
                  = \partial_\beta\frac{\partial p_0}{\partial\rho}(\partial_t\rho) 
       = -\partial_\beta\left(\frac{\partial p_0}{\partial\rho}\partial_\alpha(\rho u_\alpha)\right)
\end{equation}
\begin{equation} \label{eq:AC2a}
  {\bf AC2_\beta} = -\partial_\alpha[u_\beta\partial_\gamma(P_{\alpha\gamma} + \rho u_\alpha u_\gamma)]
\end{equation}
Neglected furthermore these terms in equation \eqref{eq:AC2a}
\begin{align*}
  \partial_\alpha\partial_\gamma(\kappa\rho\Delta\rho) &\approx 0 \\
  \partial_\alpha\partial_\gamma\left(\frac\kappa 2|\nabla\rho|^2\right) &\approx 0 \\
  \partial_\alpha\partial_\gamma\left(\kappa\partial_\theta\rho\partial_\delta\rho\right) &\approx 0 \\
  \partial_\alpha\partial_\gamma\F &\approx 0
\end{align*}
we have
\begin{equation} \label{eq:AC2b}
  {\bf AC2_\beta} = -\partial_\alpha[u_\beta\partial_\alpha p_0 
                    - u_\beta\partial_\gamma(\rho u_\alpha u_\gamma)]
\end{equation}
The term ${\bf AC_3}$ can be computed by using
\begin{equation} \label{eq:AC3a}
  \partial_t(\rho u_\beta) = u_\beta\partial_t\rho+\rho\partial_t u_\beta =
    \rho\partial_t u_\beta - u_\alpha\partial_\gamma(\rho u_\gamma)
\end{equation}
So
\begin{equation} \label{eq:AC3b}
  \rho\partial_t u_\alpha = -\partial_\beta(P_{\alpha\beta} + \rho u_\alpha u_\beta)+ 
       u_\alpha\partial_\beta(\rho u_\alpha)
\end{equation}
Therefore
\begin{align} \label{eq:AC3c}
  {\bf AC3_\beta} =& \partial_\alpha[u_\alpha(\rho\partial_t u_\beta)]  \nonumber \\
       =&-\partial_\alpha\{u_\alpha[\partial_\gamma(P_{\gamma\beta} + \rho u_\gamma u_\beta)
         +u_\beta\partial_\gamma(\rho u_\gamma)]\} \nonumber \\
       =&-\partial_\alpha\{u_\alpha\partial_\beta p_0 - \rho u_\alpha u_\gamma\partial_\gamma u_\beta]
\end{align}
Substituting equations \eqref{eq:AC1a}, \eqref{eq:AC2b} and \eqref{eq:AC3c} back into (\ref{eq:A+C}) gives
\begin{align}
  {\bf\A_\beta}+{\bf\C_\beta} =& 
         -\partial_\beta\left(\frac{\partial p_0}{\partial\rho}\partial_\alpha(\rho u_\alpha)\right) 
          \nonumber \\
        &-\partial_\alpha[u_\beta\partial_\alpha p_0+u_\alpha\partial_\beta p_0
                    - u_\beta\partial_\gamma(\rho u_\alpha u_\gamma)
                    -\rho u_\alpha u_\gamma\partial_\gamma u_\beta]
\end{align}
\begin{align}\label{eq:A+C4}
  =& -\partial_\beta\left(\frac{\partial p_0}{\partial\rho}\partial_\alpha(\rho u_\alpha)\right) 
         -\partial_\alpha(u_\beta\partial_\alpha p_0+u_\alpha\partial_\beta p_0)
         + \partial_\alpha\partial_\gamma(\rho u_\alpha u_\beta u_\gamma)\nonumber \\
  =& -\partial_\beta\left(\frac{\partial p_0}{\partial\rho}\partial_\alpha(\rho u_\alpha)\right) 
         -\partial_\alpha\left[\frac{\partial p_0}{\partial\rho}
          (u_\beta\partial_\alpha\rho+u_\alpha\partial_\beta\rho)\right]
         + \partial_\alpha\partial_\gamma(\rho u_\alpha u_\beta u_\gamma)
\end{align}
The term ${\bf\B_\beta}$ in the equation
(\ref{eq:momentum0}) can be computed by using the equilibrium 
distribution function (\ref{eq:basic_equilibrium})
\begin{align}
  \partial_\alpha\partial_\gamma\left(\sum_i w_i f_i^{eq}e_{i\alpha}e_{i\beta}e_{i\gamma}\right) =&
  \partial_\alpha\partial_\gamma\left(\sum_i w_i Bu_{\alpha}e_{i\alpha}e_{i\beta}e_{i\gamma}e_{i\theta}\right) 
  \nonumber \\
\label{eq:B0}
  =& \partial_\alpha\partial_\gamma B
\Upsilon^{(4)}(u_\gamma\delta_{\alpha\beta}+
      u_\beta\delta_{\alpha\gamma}+u_\alpha\delta_{\beta\gamma})
\end{align}
In eqn.(\ref{eq:B0}), the only free index is $\beta$, while $\alpha$ and $\gamma$ are summed over. The value of $B$ given by Swift et al
is $\rho/\Upsilon^{(2)}$\footnote{
The expression given by Swift et al is for the two-dimensional
hexagonal lattice, for which $\Upsilon^{(2)}=3c^2$,
$\Upsilon^{(4)}=3c^4/4$ and hence $B\Upsilon^{(4)}=\rho c^2/4$.}
and therefore
\begin{equation}\label{eq:B}
  {\bf\B_\beta}=\frac{\Upsilon^{(4)}}
	{\Upsilon^{(2)}}\left(
	2\partial_\beta\partial_\alpha(\rho u_\alpha)+
           \Delta(\rho u_\beta)\right)
\end{equation}
Substituting now equations (\ref{eq:A+C4}) and (\ref{eq:B}) into (\ref{eq:momentum1}) we have
\begin{align}
  \partial_t(\rho u_\beta)+\partial_\alpha(\rho u_\alpha u_\beta) =&
  -\partial_\beta p_0+\kappa\partial_\beta(\rho\Delta\rho)+\frac\kappa 2\partial_\beta
  (|\nabla\rho|^2) \nonumber
\end{align}
\begin{align}\label{eq:momentum-res1}
  -&\kappa\partial_\alpha(\partial_\alpha\rho\partial_\beta\rho)-\partial_\alpha\F_{\alpha\beta}
  +\Delta\underbrace{\left(\frac{\delta_t(2\tau-1)}2 
\frac{\Upsilon^{(4)}}{\Upsilon^{(2)}}\rho{\bf u}\right)}_{\nu\rho{\bf u}}
  \nonumber \\
  +&\partial_\beta\underbrace{\left[\delta_t\left(\tau-\frac 1 2\right)
    \left(2\frac{\Upsilon^{(4)}}{\Upsilon^{(2)}}\partial_\alpha(\rho u_\alpha)-\frac{\partial p_0}{\partial\rho}
    \partial_\alpha(\rho u_\alpha)\right)\right]}_{\lambda(\rho)
    \partial_\alpha(\rho u_\alpha)}\nonumber \\
  -&\delta_t\left(\tau-\frac 1 2\right)\partial_\alpha\left [\frac{\partial p_0}{\partial\rho}
  \left(u_\beta\partial_\alpha\rho+u_\alpha\partial_\beta\rho\right)+
  \partial_\gamma(\rho u_\alpha u_\beta u_\gamma)\right ]
\end{align}
Some terms in the equation \eqref{eq:momentum-res1} can be further simplified as:
\begin{align}
  &\kappa\partial_\beta(\rho\Delta\rho)
  +\frac\kappa 2\partial_\beta\left(|\nabla\rho|^2\right)
  -\kappa\partial_\alpha(\partial_\alpha\rho\partial_\beta\rho)\nonumber \\
  &=\kappa\partial_\beta(\rho\Delta\rho)
  +\kappa\partial_\beta\rho(\partial_\alpha\partial_\beta\rho)
  -\kappa(\partial_\alpha\partial_\beta\rho)\partial_\beta\rho
  -\kappa(\partial_\beta\rho)\Delta\rho \nonumber \\
  &=\kappa\rho\partial_\beta(\Delta\rho)
\end{align}
So, the momentum equation is:
\begin{equation}
\label{eq:NS-momentum-res2}
\begin{split}
  \partial_t(\rho u_\beta) + \partial_\alpha(\rho u_\alpha u_\beta) = -\partial_\beta p_0
    + \nu\Delta(\rho u_\beta)
    +\partial_\beta[\lambda(\rho)\partial_\alpha(\rho u_\alpha)] \\
    -\delta_t\left(\tau-\frac 1 2\right)\frac{dp_0}{d\rho}\partial_\alpha
     (u_\alpha\partial_\beta\rho+u_\beta\partial_\alpha\rho)
     -\partial_\alpha\F_{\alpha\beta}+\kappa\rho\partial_\beta(\Delta\rho)
\end{split}
\end{equation}
The extra viscous terms in the momentum equation \eqref{eq:NS-momentum-res2} are not 
Galilean invariant when density gradients are present. Comparing one sees that all of
these terms can be derived from spatial derivatives acting on the pressure, i.e. term ${\bf \C_\beta}$
in \eqref{eq:momentum0}. A possible interpretation is therefore that the Swift et al. 
condition on the second moments of the distribution function, i.e. equation 
\eqref{eqn:pressuretensor}, effectively involves an evaluation of a force term in Boltzmann's
equation (compare section \ref{sec:force} below), which has to be done to better than linear
order if unphysical viscous terms are to be avoided.

\vspace{0.4cm}\noindent Some of the non-Galilean invariant terms in \eqref{eq:NS-momentum-res2}
can be removed by an additional term $\F_{\alpha\beta}$ in the pressure tensor
(equation \eqref{eqn:pressuretensor}). First, we define:
\begin{equation} 
  \xi \equiv 2\nu-\lambda = \delta_t\left(\tau-\frac 1 2\right)\frac{dp_0}{d\rho}
\end{equation}
The equation \eqref{eq:NS-momentum-res2} can then be rewritten:
\begin{equation}
\label{eq:NS-momentum-res3}
\begin{split}
  \partial_t(\rho u_\beta) + \partial_\alpha(\rho u_\alpha u_\beta) = 
         &-\partial_\beta p_0
         + \nu\partial_\alpha(\rho\partial_\alpha u_\beta)
         + \partial_\beta(\lambda\rho\partial_\alpha u_\alpha) \\
         &+ \partial_\alpha[(\nu-\xi)u_\beta\partial_\alpha\rho
         -\xi u_\alpha\partial_\beta\rho] \\
         &+\partial_\beta[\lambda u_\alpha\partial_\alpha\rho]
         +\partial_\alpha\F_{\alpha\beta}+\kappa\rho\partial_\beta(\Delta\rho)
\end{split}
\end{equation}
By a suitable choice of the term $\F$ we can eliminate two of the three terms
with one derivative of $\rho$
\begin{equation}\label{eq:FTerm}
   \F_{\alpha\beta} = \xi(u_\beta\partial_\alpha\rho - u_\alpha\partial_\beta\rho)
        -\lambda u_\gamma\partial_\gamma\rho\delta_{\alpha\beta}
\end{equation}
and with this choice 
the momentum equation is becomes
\begin{align}\label{eq:NS-momentum-res4}
  \partial_t(\rho u_\beta) + \partial_\alpha(\rho u_\alpha u_\beta) =& 
         -\partial_\beta p_0
         + \nu\partial_\alpha(\rho\partial_\alpha u_\beta)
         + \partial_\beta(\lambda\rho\partial_\alpha u_\alpha) \nonumber\\
        &+ \nu\partial_\alpha(u_\beta\partial_\alpha\rho)
         + \kappa\rho\partial_\beta(\Delta\rho)
\end{align}
With this modified model Swift et al. found qualitatively
correct dynamics of a bubble in a liquid. This should be considered an experimental fact,
difficult to explain theoretically.
\subsection{Summary}
The model of Swift et al. has some disadvantages:
\begin{itemize}
\item The model is not Galilean invariant when the density gradients are present.
      By introducing the extra term $\F$, we can remove
	 some of the non-Galilean invariant
      terms, but not all. 
\item It is difficult to introduce the force terms into this model. We have to modify the continuity
      equation with a force terms ${\bf F}$ inside the derivative of $\rho{\bf u}$. 
      \begin{equation}
         \partial_t\rho + \nabla.(\rho{\bf u}+{\bf F}) = 0
      \end{equation}
      Therefore, this method can only be used if the force term ${\bf F}$ is negligibly small.
\end{itemize}

\section{Luo model: LBE for non-ideal dense gases}
The model of Li-Shi Luo starts from Boltzmann equation for dense gases
\cite{luo:unified-98}, also known as the Enskog equation:
\begin{equation}
  \frac{\partial f}{\partial t}+\xi.\nabla f+{\bf a}.\nabla_\xi f = \Omega
\end{equation}
where ${\bf a}$ and $\Omega$ are the 
acceleration and collision operators. With the expansion of
$\Omega$ in a Taylor series, using 
a single relaxation time approximation, and assuming the 
fluid to be
isothermal and incompressible, Luo obtains the 
following equation
\begin{subequations}
\label{eq:Luo-Enskog}
\begin{align}
  \frac{\partial f}{\partial t}+\xi.\nabla f+{\bf a}.\nabla_\xi f = -\frac g \lambda
    (f-f^{eq})+J \\
  J = -f^{eq}b\rho g(\xi-{\bf u}).\nabla ln(\rho^2g)
\end{align}
\end{subequations}
where $b$ is the second virial coefficient 
of the equation of
state for the hard-sphere system, and 
$g$ is the radial distribution function.
The equilibrium distribution function is discussed
by Luo in the continuum form
\begin{equation}
  \label{eqn:int-fcont-Luo}
  f^{eq}=\frac{\rho}{(2\pi RT)^{D/2}}\exp\left[-\frac{(\xi-{\bf u})^2}{2RT}\right]
\end{equation}
but is used in the discretized form
(\ref{eq:equilibrium-distribution-function-standard}).
\\\\
Solving (\ref{eq:Luo-Enskog}) to first
order in time we have
\begin{align}
\label{eqluo:LBE}
  f(x+\xi\delta_t,\xi,t+\delta_t)-f(x,\xi,t) =
    &-\frac 1\tau\left(f(x,\xi,t)-f^{eq}(x,\xi,t)\right) \nonumber \\
    &+J(x,\xi,t)\delta_t-{\bf a}.\nabla_\xi f(x,\xi,t)\delta_t
\end{align}
with a
force term, ${\bf a}.\nabla_\xi f$, written
\cite{luo:unified-98}
\begin{equation}\label{eqluo:force}
  {\bf a}.\nabla_\xi f = -\rho\omega(\xi)\xi_T^{-2}\left[(\xi-{\bf u})+
    \xi_T^{-2}(\xi.{\bf u})\xi\right].{\bf a}
\end{equation}
The auxiliary variable
$\xi_T$ is equal to $\sqrt{R T}$ ($\frac{c}{\sqrt{3}}$
in the $D2Q9$ model), and the weight
weight $\omega(\xi)$ denotes the weight $w_i$ for
the momentum $\xi=e_i$.
The force term $J$ is proportional to
the gradient of the function
of density $\rho^2 g$, which as in the Swift model is
evaluated by discrete differences over lattice
points. The interpretation is also that of a potential
term.
Equations (\ref{eqluo:LBE}) and (\ref{eqluo:force}) 
together with 
(\ref{eq:equilibrium-distribution-function-standard}),
(\ref{eq:macro-conservation}),
(\ref{eq:macro-momentum}), and a choice of a lattice,
define Luo's 
Lattice Boltzmann model.
\\\\
To proceed further we want to evaluate
moments with respect to momentum
of the two ``forces'' (the interior force $J$,
and the exterior force through the acceleration $a$).
The forcing term of 
eqn.\eqref{eqluo:force} is
\begin{equation}
  F_i = -w_i\rho\left[\frac{1}{\Upsilon^{(2)}}
	(e_i-{\bf u})+\frac{(e_i.{\bf u})}{\Upsilon^{(4)}
}e_i\right].{\bf a}
\end{equation}
and satisfies the following constraints:
\begin{subequations}\label{eqluo:sumFg}
\begin{align}
\label{eqluo:sumF}
  \sum_iF_i & = 0 \\
\label{eqluo:sumFe}
  \sum_i(F_ie_{i\alpha}) &=-\frac{\rho}{\Upsilon^{(2)}}
\sum_i(w_ie_{i\alpha}e_{i\beta}).{\bf a}
             =-\rho a_\alpha \\
\label{eqluo:sumFee}
  \sum_i(F_ie_{i\alpha}e_{i\beta}) &=
              -\left[-\frac{\rho u_\gamma}{\Upsilon^{(2)}}
\sum_i(e_{i\alpha}e_{i\beta}w_i)
              +\frac{\rho}{\Upsilon^{(4)}}\sum_i\big(({\bf e_i}.{\bf u})e_{i\alpha}e_{i\beta}e_{i\gamma}w_i\big)
              \right].{\bf a} \nonumber\\
            &=\rho(u_\alpha a_\beta-u_\beta a_\alpha)
\end{align}
\end{subequations}
The interior force 
term $J_i$ satisfies the following constraints:
\begin{subequations}\label{eqluo:sumJg}
\begin{align}
\label{eqluo:sumJ}
  \sum_iJ_i &= -b\rho g\left(\sum_i(f_i^{eq}e_{i\alpha})-\sum_i(f_i^{eq}u_\alpha)
               \right).\nabla ln(\rho^2g) \nonumber\\
            &= -b\rho g(\rho u-\rho u).\nabla ln(\rho^2g) = 0
\end{align}
\begin{align}
  \sum_i(J_i e_{i\alpha}) &=
               -b\rho g\left(\sum_i(f_i^{eq}e_{i\alpha}e_{i\beta})
               -\sum_i(f_i^{eq}e_{i\alpha})u_\beta\right)
	\partial_{\beta} \log (\rho^2g) \nonumber\\
\label{eqluo:sumJe}
            &= -b \Upsilon^{(2)}\partial_\alpha(\rho^2g)
\end{align}
\begin{align}
  \sum_i(J_i e_{i\alpha}e_{i\beta}) &= 
      -b\rho g\left(\sum_i(f_i^{eq}e_{i\alpha}e_{i\beta}e_{i\gamma})
      -\sum_i(f_i^{eq}e_{i\alpha}e_{i\beta})u_\beta\right).\nabla ln(\rho^2g) \nonumber\\
\label{eqluo:sumJee}
    &= b[u_\alpha u_\beta {\bf u}.\nabla - 
\Upsilon^{(2)}(u_\alpha\partial_\beta+u_\beta\partial_\alpha)]
      (\rho^2g)
\end{align}
\end{subequations}

\subsection{Chapman-Enskog analysis}\label{sec:Luo-CE}
In this section we will derive the macroscopic equations with the Chapman-Enskog
method. The earlier presented method of successive approximations can be considered
a special case of the more general and more flexible Chapman-Enskog method.
First, we apply the Taylor expansion to equation 
(\ref{eqluo:LBE}) 
\begin{equation} \label{eqluo:chapmann1}
  \sum_{n=1}^\infty\frac{\epsilon^n}{n!}\D_t^n f_i({\bf x}, t) =
  -\frac g\tau\big(f_i({\bf x},t)- f_i^{eq}({\bf x},t)\big)+J_i\epsilon-F_i\epsilon
\end{equation}
where $\epsilon=\delta_t$ and $\D\equiv(\partial_t+e_{i\alpha}\partial_\alpha)$.
We now introduce the following expansion:
\begin{subequations}
\begin{align}
  f_i=&\sum_{n=0}^\infty\epsilon^nf_i^{(n)} \\
  \partial_t =& \sum_{n=0}^\infty\epsilon^n\partial_{nt} \\
\label{eqluo:spatial1}
  \partial_\alpha =& \partial_{0\alpha}
\end{align}
\end{subequations}
The meaning of the expansion of $f$ in powers of $\epsilon$ is clear. The meaning of the formal expansion of the time derivatives in derivatives with respect to different ``times'' is that $f$ is assumed to have support in the frequency domain on bands that are separated by powers of $\epsilon$. We will look for an equation involving a ``slow time'', that is low-frequency components of $f$. Such components are Fourier transforms of $f$, where sufficienty fast oscillations are averaged out. A rigorous theory of these procedures can be given in homogenization theory \cite{Piretal}. In general one can also assume there to be fast and slow spatial variables (small and large scales).
For the present problem it is however enough to assume one spatial scale, hence \eqref{eqluo:spatial1}
\\\\
With $\epsilon^2$ order accuracy, equation \eqref{eqluo:chapmann1}
can be rewritten:
\begin{align}
  \label{eqluo:chapmann2}
  f_i\equiv f_i^{(0)}+\epsilon f_i^{(1)}+\epsilon^2 f_i^{(2)}=f_i^{eq}-
    \frac\tau g\left(\epsilon\D_t+
    \frac{\epsilon^2}{2}\D_t^2\right)f_i+\frac\tau gJ_i\epsilon-\frac\tau gF_i\epsilon
\end{align}
Group the right-hand side of the eqn.\eqref{eqluo:chapmann2} by the order of $\epsilon$.
\begin{align}
  \label{eqluo:chapmann3}
  f_i^{(0)}+\epsilon f_i^{(1)}+\epsilon^2 f_i^{(2)}=
      &f_i^{eq}-\epsilon\frac\tau g\D_t\big(f_i^{(0)}+\epsilon f_i^{(1)}+\epsilon^2 f_i^{(2)}\big)
         \nonumber \\
      &-\epsilon^2\frac{\tau}{2g}\D_t^2
         \big(f_i^{(0)}+\epsilon f_i^{(1)}+\epsilon^2 f_i^{(2)}\big) \nonumber \\
      &+\epsilon\frac\tau gJ_i-\epsilon\frac\tau gF_i
\end{align}
To successive orders in
$\epsilon$ 
we find:
\begin{subequations}\label{eqluo:dis-fg}
\begin{align}
  \label{eqluo:dis-f0}
  \epsilon^0:& \hspace{1cm} f_i^{(0)}=f_i^{eq} \\
  \label{eqluo:dis-f1}
  \epsilon^1:& \hspace{1cm} f_i^{(1)}=-\frac\tau g\D_{t_0}f_i^{(0)}+\frac\tau gJ_i- \frac\tau gF_i\\
  \label{eqluo:dis-f2}
  \epsilon^2:& \hspace{1cm} f_i^{(2)}=-\frac{\tau}{2g}\D_{t_0}^2f_i^{(0)}-
          \frac\tau g\partial_{t_1}f_i^{(0)}-\frac\tau g\D_{t_0}f_i^{(1)}
\end{align}
\end{subequations}
The goal of the calculation is to project on the hydrodynamic
modes of density and momentum. Equation (\ref{eqluo:dis-f0})
shows that the zeroth order distribution function agrees
with the equilbrium distribution function, which shares
the same density and momentum. The higher order
distribution functions therefore obey
the following constraints:
\begin{eqnarray}\label{eqluo:sum-fng}
  \sum_i f_i^{(n)} &=& 0 \qquad(n>0) \\
\label{eqluo:sum-fne}
  \sum_i (f_i^{(n)}e_{i\alpha}) &=&
	0 \qquad (n> 0)
\end{eqnarray}
Summing equation \eqref{eqluo:dis-f1} over $i$, using equations \eqref{eqluo:sumF}, and 
\eqref{eqluo:sumJ}, gives

\begin{equation}\label{eqluo:continum1}
  \partial_{t_0}\rho+\partial_{0\alpha}(\rho u_\alpha) = 0
\end{equation}
To second order in 
$\epsilon$, we sum equation \eqref{eqluo:dis-f2}
over $i$, substituting the first order solution  \eqref{eqluo:dis-f1}
which gives
\begin{equation}
  0=-\frac 1 2\left[\sum_i\D_{t_0}J_i-\sum_i\D_{t_0}F_i\right]
    +\left(\frac 1 2-\frac g\tau\right)\sum_i\D_{t_0}f_i^{(1)}-\partial_{t_1}\sum_if_i^{(0)}
\end{equation}
or
\begin{equation}
  \partial_{t_1}\rho = -\frac 1 2\partial_\alpha\left(\rho a_\alpha -
     b\Upsilon^{(2)}\partial_\alpha(\rho^2g)\right)
\end{equation}
Combining the first and the second order results for $f_i$ by
$\partial_t=\partial_{t_0}+\epsilon\partial_{t_1}$ and recalling that $\epsilon=\delta_t$,
we have a continuity equation with an (unphysical)
density diffusion term:
\footnote{The author of \cite{luo:unified-98}
claims there is no such density diffusion term,
but we find that it is there.}

\begin{align}\label{eqluo:continuity}
  \partial_t\rho + \partial_\alpha(\rho u_\alpha) =  
           -\frac{\delta_t}{2}\partial_\alpha\left(\rho a_\alpha 
           -b\Upsilon^{(2)}\partial_\alpha(\rho^2g)\right)
\end{align}
The first order of the
momentum equation can be produced by multiplying equation \eqref{eqluo:dis-f1} with
$e_{i\beta}$, then summing over $i$:
\begin{align*}
  0 &= \sum_i\D_{t_0}(f_i^{(0)}e_{i\beta}) + \sum_i(J_ie_{i\beta})- \sum_i(F_ie_{i\beta}) 
\end{align*}
By using equations \eqref{eqluo:sum-fng}, \eqref{eqluo:sumJe}, 
and \eqref{eqluo:sumFe}, we have
\begin{align}\label{eqluo:momentum0}
  \partial_{t_0}(\rho u_\beta) + \partial_\alpha(\rho u_\alpha u_\beta) 
     &= -\Upsilon^{(2)}\partial_\beta\rho-
b\Upsilon^{(2)}\partial_\beta(\rho^2g) +\rho a_\beta \nonumber\\
     &=-\partial_\beta P+\rho a_\beta
\end{align}
where 
\begin{align*}
  P=\Upsilon^{(2)}\rho(1+b\rho g)
\end{align*}
is a non-ideal pressure. Equation \eqref{eqluo:momentum0}
is hence Euler's equation.
The second order momentum equation is
\begin{align*}
  0 &=  \partial_{t_1}\sum_if_i^{(0)}e_{i\beta}
       +\frac 1 2\D_{t_0}\sum_i(J_ie_{i\beta}-F_ie_{i\beta})
       +\left(1-\frac{g}{2\tau}\right)\D_{t_0}\sum_i(f_i^{(1)}e_{i\beta})
\end{align*}
Subtituting the equations \eqref{eqluo:sum-fne}, \eqref{eqluo:sumFg} 
and \eqref{eqluo:sumJg} into above we have
\begin{equation}\label{eqluo:momentum1}
  \partial_{t_1}(\rho u_\beta) = 
      \left(\frac{g}{2\tau}-1\right)\partial_\alpha\Pi_{\alpha\beta}^{(1)}-
      \frac 1 2\D_{t_0}\sum_i(J_ie_{i\beta}-F_ie_{i\beta})
\end{equation}
where the first-order momentum flux tensor 
$\Pi_{\alpha\beta}^{(1)}$ is $\sum_if_i^{(1)}e_{i\alpha}e_{i\beta}$
and therefore
\begin{align}
  \partial_\alpha\Pi_{\alpha\beta}^{(1)} 
     &=\partial_\alpha\sum_if_i^{(1)}e_{i\alpha}e_{i\beta} \nonumber \\
     &=-\frac{\tau}{g}\partial_\alpha\sum_i\D_{t_0}\underbrace{f_i^{(0)}e_{i\alpha}e_{i\beta}}
       _{\Pi_{\alpha\beta}^{(0)}}
       +\frac{\tau}{g}\partial_\alpha\sum_i(J_ie_{i\alpha}e_{i\beta}-F_ie_{i\alpha}e_{i\beta})
\end{align}
where $\Pi_{\alpha\beta}^{(0)}$ is the zeroth-order momentum flux tensor.
Furthermore
\begin{align}\label{eqluo:Pi0}
  \D_{t_0}\Pi_{\alpha\beta}^{(0)}
      =& \partial_{t_0}\left(\Upsilon^{(2)}\rho\delta_{\alpha\beta}+\rho u_\alpha u_\beta\right)+
\Upsilon^{(2)}
\big[\partial_\alpha(\rho u_\alpha)\delta_{\alpha\beta}+
                           \partial_\alpha(\rho u_\beta)+
                           \partial_\beta(\rho u_\alpha)\big]
         \nonumber \\
      =& \Upsilon^{(2)}
\big[\partial_{t_0}\rho+\partial_\alpha(\rho u_\alpha)\big]\delta_{\alpha\beta}+
         \partial_{t_0}(\rho u_\alpha u_\beta)+
\Upsilon^{(2)}
\big[\partial_\alpha(\rho u_\beta)+\partial_\beta(\rho u_\alpha)\big]
         \nonumber \\
      =& \partial_{t_0}(\rho u_\alpha u_\beta)+
\Upsilon^{(2)}
\big[\partial_\alpha(\rho u_\beta)+\partial_\beta(\rho u_\alpha)\big]+
         \Order(\delta_t)
\end{align}
Substituting back the above results into \eqref{eqluo:momentum1}, we have,
\begin{align}\label{eqluo:momentum2}
   \partial_{t_1}(\rho u_\beta) 
      =& \Upsilon^{(2)}\frac{2\tau-g}{2g}\partial_\alpha
         \big[\partial_\alpha(\rho u_\beta)+\partial_\beta(\rho u_\alpha)\big]+
         \frac{2\tau-g}{2g}\partial_\alpha\partial_{t_0}(\rho u_\alpha u_\beta)+
         \nonumber \\
       & \frac{2\tau-g}{2g}\partial_{t_0}\sum_i(J_ie_{i\beta}-F_ie_{i\beta})+
         \frac 1 2\partial_\alpha\sum_i(J_ie_{i\alpha}e_{i\beta}-F_ie_{i\alpha}e_{i\beta})
\end{align}
Combining the first and the second order results, equations \eqref{eqluo:momentum0} and
\eqref{eqluo:momentum2}, for $f_i$ by $\partial_t=\partial_{t_0}+\epsilon\partial_{t_1}$ 
and recalling that $\epsilon=\delta_t$, we then have the 
Navier-Stokes level
momentum equation,
\begin{align}\label{eqluo:momentum3}
  \partial_t(\rho u_\beta) + \partial_\alpha(\rho u_\alpha u_\beta) = 
     -\partial_\beta P
     +\nu\partial_\alpha\big[\partial_\alpha(\rho u_\beta)+\partial_\beta(\rho u_\alpha)\big]
     +\rho a_\beta +\mathcal{E}_\beta
\end{align}
where the kinematic viscosity is
\begin{align*}
  \nu =& \Upsilon^{(2)}\frac{2\tau-g}{2g}\delta_t
\qquad \left(\,= c^2\frac{2\tau-g}{6g}\delta_t 
\hbox{in the $D2Q9$ model}\right)
\end{align*}
and various extra viscous-order terms
are grouped together in $\mathcal{E}_\beta$:
\begin{align}
  \mathcal{E}_\beta
      =& \delta_t\frac{2\tau-g}{2g}\partial_\alpha\partial_{t_0}(\rho u_\alpha u_\beta)+
         \nonumber \\
       & \delta_t\frac{2\tau-g}{2g}\partial_{t_0}\sum_i(J_ie_{i\beta}-F_ie_{i\beta})+
         \delta_t\frac 1 2\partial_\alpha\sum_i(J_ie_{i\alpha}e_{i\beta}-F_ie_{i\alpha}e_{i\beta})
         \nonumber \\
      =& \delta_t\frac{2\tau-g}{2g}\partial_\alpha\partial_{t_0}(\rho u_\alpha u_\beta)+
         \delta_t\frac{2\tau-g}{2g}\partial_{t_0}
         \left[\frac{c^2}{3}\partial_\beta(\rho^2 g)-\rho a_\beta\right]+
         \nonumber \\
       & \delta_t\frac 1 2\partial_\alpha
         \left\{b[u_\alpha u_\beta{\bf u.\nabla}-
         \Upsilon^{(2)}
(u_\alpha\partial_\beta+u_\beta\partial_\alpha)](\rho g^2)+
         \rho(u_\alpha a_\beta+u_\beta a_\alpha)\right\}
\end{align}
This momentum equation is not Galilean invariant 
when density gradients are present, and
contains some unphysical terms. If we assume that Mach number 
is small ($M$ formally of the same order as $\epsilon$
in the expansion), 
then the extra term $\partial_\alpha\partial_{t_0}(\rho u_\alpha u_\beta)$ in $ \mathcal{E}_\beta$
can be further simplified as:
\begin{align}\label{eqluo:Ebeta1}
  \partial_\alpha\partial_{t_0}(\rho u_\alpha u_\beta)
      =& \partial_\alpha\big\{u_\alpha\big[\partial_{t_0}(\rho u_\beta)-
         u_\beta\partial_{t_0}\rho\big]+u_\beta\partial_{t_0}(\rho u_\alpha)\big\} 
         \nonumber\\
      =& \partial_\alpha\big\{-u_\alpha\partial_\gamma(\rho u_\beta u_\gamma)
         +u_\alpha u_\beta\partial_\gamma(\rho u_\gamma)
         -u_\beta\partial\gamma(\rho u_\alpha u_\gamma) \nonumber\\
      & -(u_\alpha\partial_\beta P+u_\beta\partial_\alpha P)
        +\rho(u_\beta a_\alpha+u_\alpha a_\beta)\big\}+\Order(\delta_t)
\end{align}
Terms of cubic order in Mach number
($\Order({\bf u}^3)$) 
can now be neglected, and
we simplify to
\begin{align}\label{eqluo:Ebeta2}
  \partial_\alpha\partial_{t_0}(\rho u_\alpha u_\beta)
      =& -\Upsilon^{(2)}\partial_\alpha(u_\alpha\partial_\beta\rho+u_\beta\partial_\alpha\rho)
         -b\Upsilon^{(2)}
\partial_\alpha[u_\alpha\partial_\beta(\rho^2g)+u_\beta\partial_\alpha(\rho^2g)]
         \nonumber\\
       & +\partial_\alpha[\rho(u_\beta a_\alpha+u_\alpha a_\beta)]
         +\Order(\delta_t) + \Order({\bf u}^3)
\end{align}
which allows the extra terms in the momentum equation to be
rewritten as
with
\begin{align}
  \mathcal{E}_\beta = 
      \nu\partial_{\alpha}(u_{\alpha}\partial_\beta(\rho^2g))-
      \delta_t\frac{2\tau-g}{2g}\partial_{t_0}(\rho a_\beta)+
      \delta_t\tau\partial_\alpha[\rho(u_\beta a_\alpha+u_\alpha a_\beta)]
\end{align}
\subsection{Summary}
\begin{itemize}
\item This model is correct to first order (inviscid equations),
	but incorrect to second order (viscous equations).
	The continuity equation contains density diffusion
	terms, and the momentum equation various unphysical
	terms, somewhat similar to the ones appearing in the
	model of Swift and collaborators.
\end{itemize}

\section{Intermolecular interaction model}\label{sec:force}
This model was introduced by Qian, D'Humi\`eres and
Lallemand \cite{qi:lattice-92} 
and further developped by He et al
\cite{he:noideal-98}) and Chen et al. \cite{ch:two-phase-98}.
The starting point is the Boltzmann equation with a force field ${\bf F}$ in the single relaxation 
time approximation (compare \eqref{eq:Luo-Enskog})
\begin{equation}
  \label{eqn:int-basic}
  \frac{\partial f}{\partial t}+\xi.\nabla f+{\bf F}.\nabla_\xi f = -\frac{f-f^{eq}}\lambda
\end{equation}
The force ${\bf F}$ is not specified.
In the case at hand
we want eventually to recover a diffuse interface theory,
and ${\bf F}$ should therefore include density gradients.
For different proposals, see 
\cite{he:noideal-98} and \cite{ch:two-phase-98}.
The ansatz made in \cite{qi:lattice-92} is that
the gradient $\nabla_\xi f$ can be approximated by
$\nabla_\xi f^{eq}$. Since in the
continuum we have a Maxwell-Boltzmann 
equilibrium distribution function.
\begin{equation}
  \label{eqn:int-fcont}
  f^{eq}=\frac{\rho}{(2\pi RT)^{D/2}}\exp\left[-\frac{(\xi-{\bf u})^2}{2RT}\right]
\end{equation}
we further assume $\nabla_\xi f \approx  -\frac{\xi-{\bf u}}{RT}f^{eq}$
and
obtain an approximate Boltzmann equation in the form
\begin{equation}
  \label{eq:int-evolution1}
  \frac{\partial f}{\partial t}+\xi.\nabla f =
  -\frac{f-f^{eq}}\lambda+\frac{{\bf F}.(\xi-{\bf u})}{RT}f^{eq}
\end{equation}
By discretizing on a lattice and integrating over
one time step to first order we obtain a Lattice
Boltzmann scheme as
\begin{align}
  \label{eqn:int-evolution-First}
  f_i({\bf x}+{\bf e}_i\delta_t,t+\delta_t)-f_i({\bf x},t)= \hspace{5.5cm}\nonumber \\
  -\frac{f({\bf x},t)-f^{eq}({\bf x},t)}\tau+
  \frac{{\bf F}.({\bf e}_i-{\bf u})}{\Upsilon^{(2)}}f_i^{eq}\delta_t
\end{align}
since in the isothermal models Lattice Boltzmann models
$RT=\Upsilon^{(2)}$.
Comparing we see that Luo's model only differs from
\eqref{eqn:int-evolution-First} by 
a definite choice of the force term $F$, and by
a slightly different form of the external force term in
\eqref{eqluo:force}.
\\\\
The continuity and Navier-Stokes equations will
be derived by the method of successive approximations.
We note the constraints
\begin{subequations}
\begin{align}
\label{eqn:sum_f}
  \sum_i\F_i =& \frac{F_\alpha}{\Upsilon^{(2)}}\left(\sum_i e_{i\alpha}f_i^{eq}-
                     u_\alpha\sum_if_i^{eq}\right) \nonumber \\
           =& \frac{F_\alpha}{\Upsilon^{(2)}}
	(\rho u_\alpha-u_\alpha\rho) = 0 \\
\label{eqn:sum_fe}
  \sum_ie_{i\beta}\F_i =& \frac{F_\alpha}{\Upsilon^{(2)}}\left(\sum_ie_{i\alpha}
                  e_{i\beta}f_i^{eq}-
                     u_\alpha\sum_ie_{i\beta}f_i^{eq}\right) \nonumber \\
                 =& \frac{F_\alpha}{\Upsilon^{(2)}}\left(\Upsilon^{(2)}\rho\delta_{\alpha\beta}\right)
                     = {\bf F}\rho \\
\label{eqn:sum_fee}
  \sum_ie_{i\beta}e_{i\gamma}\F_i =& \frac{F_\alpha}{\Upsilon^{(2)}}\left(
       \sum_ie_{i\alpha}e_{i\beta}e_{i\gamma}f_i^{eq}-
       u_\alpha\sum_ie_{i\beta}e_{i\gamma}f_i^{eq}\right) \nonumber \\
    =& F_\alpha\left(\frac{\Upsilon^{(4)}}{(\Upsilon^{(2)})^2}
\rho(\delta_{\alpha\beta}u_\gamma+\delta_{\alpha\gamma}u_\beta+
        \delta_{\beta\gamma}u_\alpha) - u_\gamma\rho\delta_{\alpha\beta}
        -\frac{\rho}{\Upsilon^{(2)}}u_\alpha u_\beta u_\gamma\right) 
\end{align}
\end{subequations}
For the $D2Q9$ model the last expression simplifies to
\begin{equation}
\rho F_\beta u_\gamma + \rho F_\gamma u_\beta -
        \frac{3}{c^2}({\bf F.u})\rho u_\gamma u_\beta
\end{equation}
We now sum the moments of the
collision operator
$\Omega_i\equiv -1/\tau(f_i-f_i^{eq})$
over $i$. The zeroth moment to first order reads
\begin{equation} \label{eqn:continuity1}
  \partial_t\rho+\partial_\alpha(\rho u_\alpha) = \Order(\delta_t)
\end{equation}
and the next
\begin{subequations}
\begin{align}
  =&\delta_t\left(\tau-\frac 1 2\right)\left[\partial_t\partial_t\sum_i f_i^{eq}+
                2\partial_t\partial_\alpha\sum_i e_{i\alpha}f_i^{eq}+
                \partial_\alpha\partial_\beta\sum_i e_{i\alpha}e_{i\beta}f_i^{eq}\right]-\nonumber\\
   &\delta_t\tau\left(\partial_t\sum_i\F_i+\partial_\alpha\sum_i\F_i e_{i\alpha}\right)
\end{align}
\begin{align} \label{eqn:odt1}
  =&\delta_t\left(\tau-\frac 1 2\right)\left[
                \partial_t\left(\partial_t\rho+\partial_\alpha(\rho u_\alpha)\right)
                +\partial_\alpha\left(\partial_t(\rho u_\alpha)
                +\partial_\beta\sum_i e_{i\alpha}e_{i\beta}f_i^{eq}\right)\right]-\nonumber\\
   &\delta_t\tau\partial_\alpha(\rho F_\alpha)
\end{align}
\end{subequations}
Substituting in eqn.(\ref{eqn:odt1}), eqn.(\ref{eqn:continuity1}) and (\ref{eqn:Euler}) 
derived below, we see that,
\begin{align*}
  \Order(\delta_t)=&-\delta_t\left(\tau-\frac 1 2\right)[\Order(\delta_t)+
       \Order(\delta_t)+\partial_\alpha(\rho F_\alpha)]-\delta_t\tau\partial_\alpha(\rho F_\alpha)\\
   =&\Order(\delta_t^2)-\frac{\delta_t}{2}\partial_\alpha(\rho F_\alpha)
\end{align*}
with this, the continuity equation becomes,
\begin{equation} \label{eqn:continuity}
  \partial_t\rho+\partial_\alpha(\rho u_\alpha) = -\frac{\delta_t}{2} \partial_\alpha(\rho F_\alpha)+
      \Order(\delta_t^2)
\end{equation}
To obtain the momentum equation, we multiply 
$\Omega_i$ by $e_{i\beta}$
and sum over $i$. To lowest order we get an Euler equation
\begin{align}
\label{eqn:Euler}
  0 = \partial_t\sum_i e_{i\beta}f_i^{eq}+
        \partial_\alpha\sum_i
        e_{i\alpha}e_{i\beta}f_i^{eq}-\sum_i\F_i e_{i\beta}+\Order(\delta)\nonumber\\
  \partial_t(\rho u_\beta)+\partial_\alpha(\rho u_\alpha u_\beta) = -
\partial_\beta P_0
        +\rho F_\beta+\Order(\delta)
\end{align}
where $P_0 = \Upsilon^{(2)}\rho$ is the ideal gas pressure,
$P_1$ given by $\rho F_\beta = -\partial_{\beta}P_1$ 
is a correction, and the total pressure is $P=P_0+P_1$.
We note that $P$ can contain density gradient terms by
a suitable choice of $F$.
To next order in $\delta_t$ we find
\begin{align}
  \partial_t(\rho u_\beta)+\partial_\alpha(\rho u_\alpha u_\beta) =& -
\partial_\beta P_0
        +F_\beta\rho + \delta_t\left (\tau-\frac 1 2\right )\underbrace{(\partial_t
        +e_{i\alpha}\partial_\alpha)^2\sum_if_i^{eq}e_{i\beta}}_{\A_\beta} \nonumber\\
\label{eqn:momentum1}
        &-\delta_t\tau\big[\partial_t(\rho F_\beta)+\partial_\alpha(\rho F_\beta u_\alpha+
          \rho F_\alpha u_\beta)\big] + \Order(\delta^2)
\end{align}
In equation \eqref{eqn:momentum1}, a term 
$\partial_\alpha(\rho u_\alpha u_\beta {\bf F.u})$ has been
neglected because the order of this term is $\Order({\bf u}^3)$,
and $F$ will eventually also have to be taken small.
We have
\begin{subequations}
\begin{align}
  \A_\beta =& \partial_t^2\sum_if_i^{eq}e_{i_\beta}+
        2\partial_\alpha\partial_t\sum_if_i^{eq}e_{i_\beta}e_{i\alpha}+
        \partial_\alpha\partial_\gamma\sum_if_i^{eq}e_{i_\beta}e_{i\alpha}e_{i\gamma} \\
     =& \partial_t^2(\rho u_\beta)+2\partial_\alpha\partial_t\left(
\Upsilon^{(2)}
\rho\delta_{\alpha\beta}
        +\rho u_\alpha u_\beta\right)
        +\frac{\Upsilon^{(4)}}{\Upsilon^{(2)}}
\partial_\alpha\partial_\gamma(\Delta^{(4)}_{\alpha\beta\gamma\theta}u_\theta) \\
     =& \partial_t[\partial_t(\rho u_\beta)+\partial_\alpha(\rho u_\alpha u_\beta)]+
        2\Upsilon^{(2)}
\partial_{\beta}[\partial_t\rho+\partial_{\alpha}(\rho u_{\alpha})]
+ \nonumber \\
\label{eqn:termA0}
      & \partial_\alpha\partial_t(\rho u_\alpha u_\beta)+
        \frac{\Upsilon^{(4)}}{\Upsilon^{(2)}}
\partial_\alpha\partial_\alpha(\rho u_\beta)
+ 2\left(\frac{\Upsilon^{(4)}}{\Upsilon^{(2)}}
- \Upsilon^{(2)}\right)\partial^2_{\beta\alpha}(\rho u_{\alpha})
\end{align}
For the $D2Q9$ model $\Upsilon^{(4)}=(\Upsilon^{(2)})^2$,
and the last parenthesis in above vanishes.
By substituting the continuity equation \eqref{eqn:continuity1} and momentum equation 
\eqref{eqn:Euler}, with the accuracy to order of $\delta_t$, 
into \eqref{eqn:termA0} we have
\begin{align}\label{eqn:termA1}
  \A_\beta =& -\Upsilon^{(2)}\partial_t\partial_\beta\rho+\partial_t(\rho F_\beta)+
      \partial_\alpha\partial_t(\rho u_\alpha u_\beta)+
\frac{\Upsilon^{(4)}}{\Upsilon^{(2)}}
\partial_\alpha\partial_\alpha(\rho u_\beta)+\\
     =&2\left(\frac{\Upsilon^{(4)}}{\Upsilon^{(2)}}
- \Upsilon^{(2)}\right)\partial^2_{\beta\alpha}(\rho u_{\alpha})
       +\Order(\delta_t)
\end{align}
\end{subequations}
Substituting $\A_\beta$ into \eqref{eqn:momentum1}
we have the momentum equation. For the $D2Q9$ model,
as used by \cite{qi:lattice-92,he:noideal-98,ch:two-phase-98},
we have some simplifications, and finally 
\begin{align}
\partial_t(\rho u_\beta)+\partial_\alpha(\rho u_\alpha u_\beta) = -\frac{c^2}{3}\partial_\beta\rho
        + \nu\partial_\alpha\big[\partial_\beta(\rho u_\alpha)+\partial_\alpha(\rho u_\beta)\big]
        +F_\beta\rho + \nonumber \\
\label{eqn:momentum2}
       \delta_t\left(-\frac{1}{2}\partial_t(\rho F_\beta)+
	\left (\tau-\frac 1 2\right)
         \partial_\alpha\partial_t(\rho u_\alpha u_\beta)-
         \tau\partial_\alpha(\rho F_\beta u_\alpha+\rho F_\alpha u_\beta)
	\right)
\end{align}
with
\begin{align*}
  \nu = \frac{c^2\delta_t}{3}\left (\tau-\frac 1 2\right) 
      = \frac{\delta_x^2}{\delta_t}\left (\tau-\frac 1 2\right)
\end{align*}
The unwanted terms in \eqref{eqn:momentum2} are those of the
second row, all of viscous order.
Mach number $M$ is the ratio of fluid velocity and speed
of sound, which in Lattice Boltzmann models is proportional
to lattice spacing $c$. In the method of successive
approximations we have no expansion parameter $\epsilon$
with which to scale Mach number, but we can collect the
terms in   
$\A_\beta$ with smallest Mach number, and therefore neglect
the term in \eqref{eqn:momentum2}
proportional to $(\tau-\frac 1 2)$.
The momentum equations up to viscous terms hence
read
\begin{align}
\label{eqn:momentum}
\partial_t(\rho u_\beta)+\partial_\alpha(\rho u_\alpha u_\beta) = -\frac{c^2}{3}\partial_\beta\rho
        + \nu\partial_\alpha(\rho\partial_\beta u_\alpha+\rho\partial_\alpha u_\beta)
        +F_\beta\rho + \nonumber \\
         \frac{\delta_t}{2}\left(-\partial_t(\rho F_\beta)-
         \tau\partial_\alpha(\rho F_\beta u_\alpha+\rho F_\alpha u_\beta)
	\right)
        +\Order(\delta_t^2)+\Order(M^3)
\end{align}
\subsection{Summary}
\begin{itemize}
\item This model is Galliean consistent with small Mach number only.
\item  Extra terms involving the force
 appear in the continuity and momentum equations.
To have correct Navier-Stokes equations we therefore have
to assume that 
${\bf F}$ is small, that is that we are close to the ideal gas.
\end{itemize}

\section{An LBE scheme with high order accuracy}
\label{s:high-order}
The purpose of this section is to show that it suffices
to slightly modify the Lattice Boltzmann
model of section \ref{sec:force}
to get correct viscous order equations.
The model which we will describe was introduced
in \cite{he:noideal-98} with a specific choice for the
force term $\F$. 
In \cite{he:noideal-98} it is not
stated very explicitly that these modifications solve the problem
of unphysical viscous-order terms, and we therefore 
belabour this point here.
\\\\ 
We assume an isothermal situation.
The starting point is to
integrate equation   \eqref{eq:int-evolution1}
to better
than linear order for $\F$ 
\begin{align}\label{eq:int-evolution3}
  f_i({\bf x}+{\bf e}_i\delta_t,t+\delta_t)-f_i({\bf x},t)= \hspace{4.5cm}\nonumber \\
  -\int_t^{t+\delta_t}\frac{f_i-f_i^{eq}}\lambda dt+\int_t^{t+\delta_t}
  \frac{{\bf F}.({\bf e}_i-{\bf u})}{RT}f_i^{eq}dt
\end{align}
Applying the trapezoidal rule to the second integral of right-hand side
(but not to the first), we have the
Lattice Boltzmann model
(compare
\eqref{eqn:int-evolution-First})
\begin{align}
  \label{eq:int-evolution4}
  f_i({\bf x}+{\bf e}_i\delta_t,t+\delta_t)-f_i({\bf x},t)= \hspace{4.5cm}\nonumber \\
  -\frac{f_i-f_i^{eq}}\tau \Big |_t+
  \frac{\delta_t}{2}\biggr [\frac{{\bf F}.({\bf e}_i-{\bf u})}{RT}f_i^{eq}\Big |_{t+\delta_t}+
  \frac{{\bf F}.({\bf e}_i-{\bf u})}{RT}f_i^{eq}\Big |_t\biggr ]
\end{align}
where $\tau=\lambda/\delta_t$ is the non-dimensional relaxation time. The right-hand side
involves the quantities evaluated at $t+\delta_t$. To eliminate this implicitness, we introduce the new variable:
\begin{equation}
	\label{eq:hi-def}
  h_i=f_i-\frac{F.({\bf e}_i-u)}{2RT}f_i^{eq}\delta_t
\end{equation}
in term of which the Lattice Boltzmann equation \eqref{eq:int-evolution4} is:
\begin{align}
  \label{eqn:int-evolution}
  h_i({\bf x}+{\bf e}_i\delta_t,t+\delta_t)-h_i({\bf x},t)= \hspace{5.5cm}\nonumber \\
  -\frac{h({\bf x},t)- h^{eq}({\bf x},t)}\tau+
  \frac{{\bf F}.({\bf e}_i-{\bf u})}{RT}f_i^{eq}\delta_t
\end{align}
Another interpretation of this procedure is that we evaluated all terms in equation 
\eqref{eq:int-evolution4} to the same order, with a collision operator for the original variables
$f_i$ that was actually not of the single relaxation time form.
Instead it was something different
that gave the single relaxation time form for the
auxiliary variables $h_i$.
\\\\
The equilibrium distribution for $h$ contains
terms up to three velocities, and reads
\begin{equation}
	\label{eq:hi-eq-def}
  h_i^{eq}=\left[1-3\frac{F.({\bf e}_i-u)}{2c^2}\delta_t\right]f_i^{eq}
\end{equation}
where $f_i^{eq}$ is the standard choice of 
\eqref{eq:equilibrium-distribution-function-standard}.
The macroscopic variables given by eqn.(\ref{eq:hi-def}) become
\begin{align}
  \rho =& \sum_i h_i \\
  \rho{\bf u} =& \sum_i h_i{\bf e}_i +\frac 1 2\rho{\bf F}\delta_t
\end{align}
\\\\
We now wish to compute the macroscopic equations
by Chapman-Enskog expansions.
We will for notational
ease use both $h_i$ and $f_i$ in the expansions.
Since these
two quantities differ by a term to order $\epsilon$
(i.e. equations \eqref{eq:hi-def} and \eqref{eq:hi-eq-def}),
a choice has to be made for which of the two in
which we count powers of $\epsilon$.
The natural choice is to count in orders of $\epsilon$
for $f_i$, but to perform the expansion in $h_i$,
since the equations for $h_i$ are explicit.
This will mix orders for $h_i$, but in an unambiguous
manner, as the expansion can be mapped back to order
by order in $\epsilon$ for $f_i$. To
$\epsilon^2$ order accuracy we have
\begin{align}
  \label{eqn:int-chapmann02}
  h_i\equiv h_i^{(0)}+\epsilon h_i^{(1)}+\epsilon^2 h_i^{(2)}=h_i^{eq}-
    \tau\left(\epsilon\D_t+
    \frac{\epsilon^2}{2}\D_t^2\right)h_i+\epsilon\tau\F_i
\end{align}
where each term $h_i^{(n)}$ includes a term of order $\epsilon$,
and where  
\begin{align*}
  \F_i = \frac{{\bf F}.({\bf e}_i-{\bf u})}{RT}f_i^{eq}
\end{align*}
Regrouping the right-hand side of eqn.(\ref{eqn:int-chapmann02})
 by the leading order of $\epsilon$ we have
\begin{align}
  \label{eqn:int-chapmann03}
  h_i^{(0)}+\epsilon h_i^{(1)}+\epsilon^2 h_i^{(2)}=
      &h_i^{eq}-\epsilon\tau\D_t\big(h_i^{(0)}+\epsilon h_i^{(1)}+\epsilon^2 h_i^{(2)}\big)
         \nonumber \\
      -&\epsilon^2\frac{\tau}{2}\D_t^2
         \big(h_i^{(0)}+\epsilon h_i^{(1)}+\epsilon^2 h_i^{(2)}\big) \nonumber \\
      +&\epsilon\tau\F_i
\end{align}
which is
\begin{subequations}
\begin{align}
  \label{eqn:dis-h0}
  \epsilon^0:& \hspace{1cm} h_i^{(0)}=h_i^{eq} \\
  \label{eqn:dis-h1}
  \epsilon^1:& \hspace{1cm} h_i^{(1)}=-\tau\D_{t_0}h_i^{(0)}+\tau\F_i \\
  \label{eqn:dis-h2}
  \epsilon^2:& \hspace{1cm} h_i^{(2)}=-\frac\tau 2\D_{t_0}^2h_i^{(0)}-
          \tau\partial_{t_1}h_i^{(0)}-\tau\D_{t_0}h_i^{(1)}
\end{align}
\end{subequations}
Equation \eqref{eqn:dis-h0} here determines $h_i^{(0)}$
to be of mixed zeroth and first order in $\epsilon$,
i.e. given by \eqref{eq:hi-eq-def}.
Since the hydrodynamic modes belong to to equilbrium
distribution functions, we have for 
$n$ greater than zero 
the constraints
\begin{align}
  \sum_i h_i^{(n)} =& 0 \\
  \sum_i h_i^{(n)}e_{i\alpha} =& 0
\end{align}
For the lowest order $h_i^{(0)}=h_i^{eq}$ we have on the other hand
\begin{align}
  \label{eqn:sum-h}
  \sum_i h_i^{(0)} =& \sum_i\left[f_i^{eq}-\frac{\epsilon} 2\F\right] = \rho \\
  \label{eqn:sum-he}
  \sum_i h_i^{(0)}e_{i\alpha} =& \sum_i f_i^{eq}e_{i\alpha}-
           \frac{\epsilon} 2\sum_i \F_i e_{i\alpha} =
           \rho{\bf u}-\frac{\epsilon} 2\rho{\bf F}\\
  \label{eqn:sum-hee}
  \sum_i h_i^{(0)}e_{i\alpha}e_{i\beta} =& \sum_i f_i^{eq}e_{i\alpha}e_{i\beta}-
           \frac{\epsilon} 2\sum_i \F_i e_{i\alpha}e_{i\beta} \nonumber \\
           =& \Upsilon^{(2)}\rho\delta_{\alpha\beta}+\rho u_\alpha u_\beta -
              \frac{\epsilon}2\left[\rho F_\alpha u_\beta+\rho F_\beta u_\alpha
               -\frac{1}{\Upsilon^{(2)}}(F.u)\rho u_\alpha u_\beta\right]
\end{align}
Summing equation (\ref{eqn:dis-h1}), using equations (\ref{eqn:sum-h}), (\ref{eqn:sum-he})
and (\ref{eqn:sum_f}), gives
\begin{subequations}
\begin{align}
  \partial_{t_0}\rho + \partial_{\alpha}(\rho u_\alpha) =
       \frac{\epsilon}{2}\partial_{\alpha}(\rho F_\alpha)
\end{align}
\end{subequations}
To second-order in $\epsilon$ we get by summing (\ref{eqn:dis-h2})
\begin{subequations}
\begin{align}
  0 = -\frac{1}{2} \D_{t_0}\sum_i\F_i+\left(\frac{1}{2}-\tau\right)
      \D_{t_0}\sum_ih_i^{(1)}-\partial_{t_1}\sum_i h_i^{(0)}
\end{align}
The first and the third terms is substituted from equations \eqref{eqn:sum_f}, \eqref{eqn:sum_fe}
and \eqref{eqn:sum-h}. The second term is zero,
and the two others give
\begin{align}
§  \partial_{t_1}\rho &= -\frac 1 2 \partial_{\alpha}(\rho F_\alpha)
\end{align}
\end{subequations}
Combining the first and the second order results for $h_i$ by
$\partial_t=\partial_{t_0}+\epsilon\partial_{t_1}$ 
we have the continuity equation:
\begin{align}\label{eqn2:continuity}
  \partial_t\rho + \nabla.(\rho{\bf u}) = 0
\end{align}
valid up to order $\epsilon^2$. 
In other words, there is no density diffusion
term in this equation.
\\\\
To first order, the momentum equation can be computed by multiplying equation (\ref{eqn:dis-h1})
with $e_{i\beta}$, then summing over $i$, and using \eqref{eqn:sum-hee}:
\begin{subequations}
\begin{align}
  \partial_{t_0}(\rho u_\beta)+\partial_\alpha(\sum_ie_{i\alpha}e_{i\beta}h_i^{(0)}) =
      \frac{\epsilon}{2}\partial_{t_0}(\rho F_\beta)+\rho F_\beta
\end{align}
which is
\begin{align}
  \partial_{t_0}(\rho u_\beta)+\partial_\alpha(\rho u_\alpha u_\beta) =&
      -\Upsilon^{(2)}\partial_\beta\rho
      +\frac{\epsilon}2\partial_\alpha(\rho F_\alpha 
	u_\beta+\rho F_\beta u_\alpha) \nonumber\\
\label{eqn:momentum-1}
     &-\frac{\epsilon}{2\Upsilon^{(2)}}\partial_\alpha[(F.u)\rho u_\alpha u_\beta]
      +\frac{\epsilon}{2}\partial_{t_0}(\rho F_\beta)+\rho F_\beta
\end{align}
\end{subequations}
The second order momentum equation are obtained by multiplying equation \eqref{eqn:dis-h2} with
$e_{i\beta}$ and summing over $i$,
which leads to
\begin{subequations}
\begin{align}
  0 &= \partial_{t_1}(\sum_ih_i^{(0)}e_{i\beta})+\frac 1 2\D_{t_0}\sum_i\F_i e_{i\beta}
       -\left(\frac{1}{2\tau}-1\right)\D_{t_0}\sum_i(h_i^{(1)}e_{i\beta})
\end{align}
Substituting in equations \eqref{eqn:sum-he}, 
\eqref{eqn:sum_fe} and \eqref{eqn:sum_fee}, we have
\begin{align}
  \partial_{t_1}(\rho u_\beta) =& \left(\frac{1}{2\tau}-1\right)\partial_{\alpha}
                                  \sum_i(h_i^{(1)}e_{i\alpha}e_{i\beta})\nonumber\\
\label{eqn:momentum-2}
     &-\frac 1 2\partial_{t_0}(\rho F_\beta)-\frac 1 2\partial_{\alpha}\left(\rho F_\alpha u_\beta 
      + \rho F_\beta u_\alpha -\frac{1}{\Upsilon^{(2)}}({\bf F.u})\rho u_\alpha u_\beta\right)
\end{align}
\end{subequations}
Combining the first and the second order results, equations \eqref{eqn:momentum-1} and
\eqref{eqn:momentum-2}, for $h_i$ by $\partial_t=\partial_{t_0}+\epsilon\partial_{t_1}$,
we have,
\begin{align}\label{eqn:momentum-3}
  \partial_t(\rho u_\beta)+\partial_\alpha(\rho u_\alpha u_\beta) =
     -\Upsilon^{(2)}\partial_\beta\rho+\epsilon\left(\frac{1}{2\tau}-1\right)\partial_\alpha
      \Pi_{\alpha\beta}^{(1)}+\rho F_\beta
\end{align}
where $\Pi_{\alpha\beta}^{(1)}=\sum_ie_{i\alpha}e_{i\beta}h_i^{(1)}$ is the first-order momentum flux tensor.
We have
\begin{align}
  \Pi_{\alpha\beta}^{(1)} &= \sum_ie_{i\alpha}e_{i\beta}h_i^{(1)} 
             = -\tau\sum_ie_{i\alpha}e_{i\beta}\D_{t_0}h_i^{(0)}
               +\tau\sum_ie_{i\alpha}e_{i\beta}\F_i\nonumber\\
            &= -\tau\left(\partial_{t_0}\sum_ie_{i\alpha}e_{i\beta}h_i^{(0)}
               +\partial_{\gamma}\sum_ie_{i\alpha}e_{i\beta}e_{i\gamma}h_i^{(0)}\right)
               +\tau\sum_ie_{i\alpha}e_{i\beta}\F_i\nonumber
\end{align}
and in which the sums can be simplified to
\begin{subequations}\label{eqn:Pih}
\begin{align}
  \sum_ie_{i\alpha}e_{i\beta}h_i^{(0)} 
     =& \sum_ie_{i\alpha}e_{i\beta}f_i^{(0)} - 
        \frac{\epsilon}{2}\sum_ie_{i\alpha}e_{i\beta}\F_i \\
  \sum_ie_{i\alpha}e_{i\beta}e_{i\gamma}h_i^{(0)} 
     =& \sum_ie_{i\alpha}e_{i\beta}e_{i\gamma}f_i^{(0)} - 
        \frac{\epsilon}{2}\sum_ie_{i\alpha}e_{i\beta}e_{i\gamma}\F_i
\end{align}
\end{subequations}
The second terms in equations \eqref{eqn:Pih} are order of $\epsilon$. Therefore, when we 
substitute them back into \eqref{eqn:momentum-3}, 
these terms become $\epsilon^2$,
and should be neglected. 
With this simplification
$\Pi_{\alpha\beta}^{(1)}$ can be rewritten,
\begin{align}
  \Pi_{\alpha\beta}^{(1)} =  
         -\tau\underbrace{\left(\partial_{t_0}\sum_ie_{i\alpha}e_{i\beta}h_i^{(0)}
         +\partial_{\gamma}\sum_ie_{i\alpha}e_{i\beta}e_{i\gamma}h_i^{(0)}\right)}
         _{\D_{t_0}\Pi_{\alpha\beta}^{(0)}}
         +\tau\sum_ie_{i\alpha}e_{i\beta}\F_i\nonumber
\end{align}
where $\Pi_{\alpha\beta}^{(0)}$ is zeroth-order 
momentum flux tensor, and the convective
derivative has been computed above in
\eqref{eqluo:Pi0}
\begin{align}
  \D_{t_0}\Pi_{\alpha\beta}^{(0)}
     =& \partial_{t_0}(\rho u_\alpha u_\beta)+
        \Upsilon^{(2)}\big[\partial_\alpha(\rho u_\beta)+\partial_\beta(\rho u_\alpha)\big]+
        \Order(\delta_t) \\
     =& \Upsilon^{(2)}(\rho\partial_\alpha u_\beta+\rho\partial_\beta u_\alpha)+
        \rho(u_\alpha F_\beta+u_\beta F_\alpha)+
        \Order(\delta_t)+\Order({\bf u}^3)
\end{align}
The first-order momentum flux tensor
is therefore
in the small Mach number approximation and to
lowest order in $\epsilon$ equal 
to
\begin{align}
  \Pi_{\alpha\beta}^{(1)} = -\tau\Upsilon^{(2)}§(\rho\partial_\alpha u_\beta+\rho\partial_\beta u_\alpha)
\end{align}
Substituting back the above results into \eqref{eqn:momentum-3} we have the momentum equation,
\begin{align}\label{eqn:momentum-4}
  \partial_t(\rho u_\beta)+\partial_\alpha(\rho u_\alpha u_\beta) =
     -\Upsilon^{(2)}\partial_\beta\rho
     +\nu\partial_\alpha(\rho\partial_\alpha u_\beta+\rho\partial_\beta u_\alpha)
     +\rho F_\beta
\end{align}
where
\begin{align*}
  \nu = \Upsilon^{(2)}\delta_t\left (\tau-\frac 1 2\right) 
\end{align*}
The final equation is the Navier-Stokes 
equation of an ideal gas with a force field.
This force field can be an external force, but
it can also be an internal force which describes 
capillary forces and non-ideal corrections.
By comparison with 
\eqref{eq:momentum-diffuse-interface}
we see that it suffices to choose
$\rho F_{\beta}= \Upsilon^{(2)}\partial_\beta\rho +
\partial_{\alpha}T_{\alpha\beta}$ to describe
isothermal flow with reversible stress tensor
$T_{\alpha\beta}$. If we for instance wish to have the
capillary stress tensor of \eqref{eq:stress-tensor} we
should choose ${\bf F} =
\left(\Upsilon^{(2)} + (\rho(f-\mu)' - K\nabla^2\rho\right)
{\bf \nabla}\log\rho$, where the Laplacian 
and the gradient can be computed
by discretization on the grid as in
\cite{sw:lattice-95}. 

\subsection{Summary}
\begin{itemize}
  \item This modifies second-order Lattice Boltzmann equation
	gives rise to correct continuity and momentum equations
	to viscous order. The local and nonlocal presure terms
	in a diffuse interface model can be introduced via
	the force field $F$. 
\item As in previous sections, a small Mach number expansion
	has been used, such that terms with three velocities
	are ignored. If desired this can be improved by using
	lattices with higher isotropy, following
	\cite{qian-orszag,chen-ohashi-akiyama}.
\item The model is isothermal. 
\end{itemize}
\section{Thermal lattice Boltzmann}
\label{s:thermal}
\subsection{General considerationsn}
The equations of non-isentropic fluid flow 
are the continuity equation, the momentum equation,
an equation of state and either the
energy or entropy transport equations.
In the context of diffuse interface theory,
the energy and entropy transport equations
are
\eqref{eq:internal-energy-diffuse-interface}
and \eqref{eq:entropy-diffuse-interface}
above.
One difficulty in constructing a Lattice Boltzmann
scheme for non-isentropic flow is that we have
so far not defined the internal energy or the
entropy in terms of the distribution function.
\\\\
Let us start by remarking that temperature can
be defined in terms of the thermal kinetic energy
as
\begin{align}
  \rho\varepsilon_T = \frac 1 2\sum_i f_i(e_{i\alpha}-{\bf u})^2
\end{align}
where $\rho$ is the mass density, and 
$\varepsilon_T=\frac{D}{2}RT$. In the rest of this
section we will assume units have been chosen
such that the gas constant $R$ is one.
A first requirement of a nonisotropic Lattice Boltzmann
model is that temperature is not a constant, i.e.
that the equilibrium distribution function
\eqref{eq:equilibrium-distribution-function-standard}
is modified to depend on $T$.
\\\\
In equilibrium,
the internal energy and entropy densities per unit volume
are related by (\cite{ChaikinLubensky}, note that $\varepsilon_I$
and $s$ here stand for densities per unit mass)
\begin{equation}
d(\rho\varepsilon_I) = T d(\rho s) + \mu d\rho
\end{equation}
The free energy density and the pressure on the other
obey
\begin{equation}
\rho f = \rho\left(\varepsilon_I -Ts\right)
\quad p = -\rho(f-\mu)
\end{equation}
from which follows
\begin{equation}
d \varepsilon_I = -\frac{p}{\rho^2}\,d \rho  + T ds
\quad
d f = -\frac{p}{\rho^2}\,d \rho  - s dT
\end{equation}
If we with slight abuse of notation denote
the free energy functional density per unit mass
in the Cahn-Hilliard theory by 
\begin{equation}
f[\rho,\nabla\rho,T] = f(\rho,T) +\frac{K}{2\rho}|\nabla\rho|^2 
\end{equation}
where $f(\rho,T)$ is the free energy density of a homogeneous
phase at density $\rho$ and temperature $T$, then
\begin{eqnarray}
p &=& -\rho^2\,\frac{\delta f[\,\,]}{\delta \rho}|_{T}
= - \rho^2 \frac{\partial f}{\partial \rho}
- \frac{K}{2}|\nabla\rho|^2 - K\rho\nabla^2\rho \\ 
s &=& -\frac{\delta f[\,\,]}{\delta T}|_{\rho}
	=  -\frac{\partial f}{\partial T}
\end{eqnarray}
We note that $p$ is equal to the diagonal component
of the stress tensor in the second formulation
discussed in section \ref{s:multiphase}.
We also note that in equilibrium, as we assume here,
the non-diagonal terms of the stress tensor are zero,
and the stress tensor can therefore be considered
a non-equilibrium generalization of pressure. 
Entropy density in diffuse interface theory is just
a function of $\rho$ and $T$, while internal energy
density depends on density gradients, i.e.
\begin{equation}
\varepsilon_I[\rho,\nabla\rho,T] = f[\,\,] + Ts =
\varepsilon_I(\rho,s)
 +\frac{K}{2\rho}|\nabla\rho|^2 
\end{equation}
where $(\varepsilon_I, s)$ and $(f,T)$ forms
a Legendre transform pair:
\begin{equation}
\varepsilon_I(\rho,s) =
f(\rho,T) + s\,T \qquad s = - \frac{\partial f}{\partial T}
\end{equation}
Thermodynamics in a system with mass motion follows
from the free energy (see \cite{ChaikinLubensky}, section 8.4) 
\begin{equation}
F(T,V,M,{\bf v} ) = E - TS - {\bf P}\cdot {\bf v}
\end{equation} 
where ${\bf P}$ is total momentum, and ${\bf v}$
is the velocity of the rest frame of the fluid
relative to the laboratory. 
We have ${\bf P} = M{\bf v}$, and the
internal energy with total momentum ${\bf P}$ is 
\begin{equation}
E(S,V,M,{\bf P}) = E(S,V,M,0) +\frac{1}{2}\frac{P^2}{M}
\end{equation}
from which follows 
\begin{equation}
F(T,V,M,{\bf v}) =  F(T,V,M,0)-\frac{1}{2}M{\bf v}^2
\end{equation}
The internal energy $\varepsilon_I$, the free energy density
$f$ and momentum density per unit mass $\xi$\footnote
{In equilibrium the average momentum density per
unit mass is the velocity $v$.} 
in a system in motion are therefore related by
\begin{equation}
f(\rho,T,v) = \varepsilon_I(\rho,s,\xi) -Ts - {\bf v}^2 \quad
\varepsilon_I(\rho,s,\xi) =  \varepsilon_I(\rho,s,0)
+ \frac{1}{2} {\bf \xi}^2 
\end{equation}

\subsection{Non-isentropic Lattice Boltzmann}
The equilibrium distribution function can be 
generalized to
depend on temperature $T$ as\footnote{We assume 
units such that $R=1$.}
\begin{equation}
\label{eq:equilibrium-distribution-function-generalized}
  f_i^{eq} = \rho\left[1+\frac{(e_i.{\bf u})} {T}+
\frac{(e_i.{\bf u})^2\Upsilon^{(2)}}{2T\Upsilon^{(4)}} -
\frac{{\bf u}^2} {2 T}\right]
\end{equation}
We consider the continuous-time Boltzmann equation
in the form
\begin{equation}
  \label{eqn:int-generalized}
  \frac{\partial f}{\partial t}+
	\xi.\nabla f+{\bf F}.\nabla_\xi f = \Omega(f)
\end{equation}
where we identify $\xi = \partial_{\xi}\varepsilon(\rho,s,\xi)$
and  ${\bf F} = -\partial_{x}\varepsilon(\rho(x),s(x),\xi)$.
The left-hand side of \eqref{eqn:int-generalized} is
Liouville's equation.
\\\\
Here is a list of unanswered questions, that we
feel is pertinent:
\begin{center}
\begin{itemize}
\item What quantities should the collision operator $\Omega$
	conserve ? Presumably mass, momentum and energy ?
\item Can this then be done with a single relaxation time
	scheme ?
\item The internal energy density $\varepsilon_I$ is a function
	of $\rho$ and $s$, but $s$ is a function of 
	$\rho$ and $T$. Hence we can consider $\varepsilon_I$
	a function of $\rho$ and $T$, both quantities which
	are naturally defined for the LBE. More explicitly,
	on the lattice, we could have
	\begin{equation}
	\varepsilon_I(x,t) =
	\varepsilon_I^0(\rho(x,t),T(x,t))
	+ \frac{K}{2\rho(x,t)\Upsilon^{(2)}}
	\sum_i w_i (\rho(x+e_i,t)-\rho(x,t))^2
	\end{equation}
	where $\varepsilon_I^0(\rho(x,t),T(x,t))$
	stands for the bulk contribution to the internal
	energy, and the rest is a discrete approximation
	to the gradient terms.

	How to construct an LBE scheme that conserves such a beast?
\end{itemize}
\end{center}

\subsection{An attempt to generalize section 7}
The LBE models in the previous sections are 
all isothermal.
The objective of the present section is 
to describe work towards adapting
the model described in section \ref{s:high-order}
to non-isothermal flow
in a non-ideal fluid. The procedure follows the
one for heat flow in an one-phase fluid with an ideal
gas equation of state introduced by Alexander et al
\cite{alex:thermal-93}.
This procedure has a limitation, in that the Prandtl
number can not be varied.
\\\\
Several Lattice Boltzmann models of fluids with heat and
mass flow have been introduced in the literature. Most of
these share a defect of the one presented here, in that
an equation is derived for the variation of temperature.
The general equation of heat transfer
\eqref{eq:entropy-diffuse-interface} involves the
variation of entropy density, and only reduces to
an equation for temperature in the incompressible
limit, see \cite{landau} \S\,50.
We will not attempt a review
of these models, but refer to
\cite{MalevanetsKapral,McNamaraGarciaAlder,IhleKroll,chen-ohashi-akiyama,palmer-rector}. 
The work of Palmer \& Rector \cite{palmer-rector} deserves however
to be singled out, as it uses a different approach where the
Prandtl number can be changed, and simulates
the energy transport equation 
\eqref{eq:internal-energy-diffuse-interface},
albeit without the effects of viscous heating and interstitial
working.
\\\\
Let us start with defining the thermal kinetic energy
as
\begin{align}
  \rho\varepsilon = \frac 1 2\sum_i f_i(e_{i\alpha}-{\bf u})^2
\end{align}
where $\rho$ is the mass density, and 
$\varepsilon=\frac{D}{2}RT$\footnote{
As an aside, let us note that a natural definition
of entropy density per unit mass would be
\begin{align}
  \rho s = - \sum_i f_i \log f_i \nonumber
\end{align}
where in equilibrium $s$ is a function of $\rho$ and
$T$. A Lattice Boltzmann scheme starting from such
a definition would naturally lead to a 
convective derivative of entropy density, as
on the left-hand side of \eqref{eq:entropy-diffuse-interface}.
The right hand side of \eqref{eq:entropy-diffuse-interface}
however only follows if the equilibrium function
is chosen in an suitable form: this work is
not attempted here.
}.
The equilibrium distribution function 
\eqref{eq:equilibrium-distribution-function-standard}
only leads to isothermal flows, as $\varepsilon$
is then $\Upsilon^{(2)}$.
However, by changing the equilibrium function to
be a function of $\rho$, $u$ and $T$
as in \cite{alex:thermal-93}, we can have the following
set of constraints:
\begin{subequations}
\label{th:f-moments}
\begin{align}
	\label{th:f-zero-moment}
  \sum_i f_i^{eq} =& \rho \\
	\label{th:f-first-moment}
  \sum_i f_i^{eq} e_{i\alpha} =& \rho u_{\alpha} \\
	\label{th:f-second-moment}
  \sum_i f_i^{eq} 
	e_{i\alpha} e_{i\beta} =&
	\rho u_{\alpha}u_{\beta} + 
	\rho RT \delta_{\alpha\beta} \\
	\label{th:f-third-moment}
  \sum_i f_i^{eq} e_{i\alpha} e_{i\beta} e_{i\gamma} =&
	\rho u_\alpha u_\beta u_\gamma +
	\rho RT
		(\delta_{\alpha\beta}u_{\gamma} +
                         \delta_{\alpha\gamma}u_{\beta} +
                         \delta_{\beta\gamma}u_{\alpha}) \\
	\label{th:f-fourth-moment}
  \sum_i f_i^{eq} e_{i\alpha}^2 e_{i\beta} e_{i\gamma} =&
	\rho u_\alpha^2 u_\beta u_\gamma + \nonumber \\
	&(D+2)\rho (RT)^2\delta_{\beta\gamma} +
	\rho(RT) u_{\alpha}^2\delta_{\beta\gamma} +
	(D+4)\rho (RT) u_{\beta}u_{\alpha}
\end{align}
\end{subequations}
We now introduce an auxiliary quantity $h_i$ as
in section \ref{s:high-order} which obeys
\begin{subequations}
\label{th:h-moments}
\begin{align}
	\label{th:h-zero-moment}
  \sum_i h_i^{eq} =& \rho \\
	\label{th:h-first-moment}
  \sum_i h_i^{eq} e_{i\alpha} =& \rho u_{\alpha} -
	\frac{1}{2\Upsilon^{(2)}}F_\alpha\rho(RT)
\delta_t \\
	\label{th:h-second-moment}
  \sum_i h_i^{eq} 
	e_{i\alpha} e_{i\beta} =&
	\rho u_{\alpha}u_{\beta} + 
	\rho RT \delta_{\alpha\beta} -
	\frac{1}{2\Upsilon^{(2)}}(F_\beta u_\alpha+F_\alpha u_\beta)
	\rho RT \delta_t 
\end{align}
\end{subequations}
and a force term in the Lattice Boltzmann equations 
\eqref{eqn:int-evolution} which obeys
\begin{subequations}
\label{th:Fi-moments}
\begin{align}
	\label{th:Fi-zero-moment}
  \sum_i \F_i =& 0 \\
	\label{th:Fi-first-moment}
  \sum_i \F_i e_{i\beta} =& \frac{1}{\Upsilon^{(2)}}
	\rho RT F_{\beta} \\
	\label{th:Fi-second-moment}
  \sum_i \F_i 
	e_{i\beta} e_{i\gamma} =& \frac{1}{\Upsilon^{(2)}}\rho RT
	(F_{\beta}u_\gamma+F_\gamma u_\beta) \\
	\label{th:Fi-third-moment}
  \sum_i \F_i 
	e_{i\alpha}^2 e_{i\beta} =& \frac{1}{\Upsilon^{(2)}}\left(
	(D+2)\rho (RT)^2 F_\beta+
	 \rho RT u_\alpha^2 F_\beta+
	2 \rho RT F_\alpha u_\alpha u_\beta)\right)
\end{align}
\end{subequations}
We compute the macroscopic equations with Chapman-Enskog expansions
as in section \ref{s:high-order}. The constraints we use to
close the equations are the zeroth, first and second moments of
the collision operator, that is
\begin{equation}
\label{eqn:dis-h1}.
\sum_i (f_i - f^{eq}_i) 
=\sum_i (f_i - f^{eq}_i) e_{i\alpha}
=\sum_i (f_i - f^{eq}_i) e_{i\alpha} e_{i\alpha} = 0
\end{equation}
The first and second order in expansion parameter 
$\epsilon$ of the continuity equation read:
\begin{subequations}
\label{th:continuity}
\begin{align}
	\label{th:continuity0}
  \partial_{t_0}\rho + \partial_\alpha(\rho u_\alpha) &= 
	\epsilon\frac{1}{2\Upsilon^{(2)}}
	\partial_\alpha(F_\alpha\rho RT) \\
	\label{th:continuity1}
  \partial_{t_1}\rho &=-\frac{1}{2\Upsilon^{(2)}}
\partial_\alpha(F_\alpha\rho RT)
\end{align}
Combining the first and the second order results by
$\partial_t=\partial_{t_0}+\epsilon\partial_{t_1}$ 
we have the continuity equation:
\begin{align}	\label{th:continuity3}
  \partial_t\rho + \nabla.(\rho{\bf u}) = 0
\end{align}
\end{subequations}
The first-order in $\epsilon$ of the momentum equation is:
\begin{align}	\label{th:momentum0}
  \partial_{t_0}(\rho u_\beta) + \partial_\alpha(\rho u_\alpha u_\beta) = &-
	\partial_\beta P_0 +
	\epsilon\frac{1}{2\Upsilon^{(2)}}
	(F_\beta u_\alpha+F_\alpha u_\beta)\rho RT +
	\nonumber \\
	&\epsilon\frac{1}{2\Upsilon^{(2)}}
	\partial_{t_0}(F_\beta\rho RT) +
	\frac{1}{\Upsilon^{(2)}}F_\beta\rho RT
\end{align}
with the ideal part of the
pressure $P_0=\rho RT$. The first-order in $\epsilon$ of the 
temperature 
equation is calculated by taking the second moment of \eqref{eqn:dis-h1}.
\begin{subequations}
\begin{align}	\label{th:energy0}
  \partial_{t_0}&\left(\frac{D}{2}\rho RT+\frac{1}{2}\rho u_\alpha^2\right) =\nonumber\\
	&-\frac{1}{2}\partial_\beta\left(\rho u_\alpha^2 u_\beta+
	\frac{(2+D)}\rho RT u_\beta\right) +
	\frac{1}{\Upsilon^{(2)}}\rho RT F_\beta u_\beta +
	\frac{\epsilon}{4}\D_{t_0}\sum_i\F_ie_{i\alpha}^2
\end{align}
\end{subequations}
To second-order, the temperature
equation can be computed by the same way from equation
\eqref{eqn:dis-h2}. 
\begin{subequations}
\begin{align}	\label{th:energy1}
  \partial_{t_1}&\left(\frac{D}{2} \rho RT+\frac{1}{2}\rho u_\alpha^2\right)=\nonumber\\
	&\frac1 2\left(\frac{1}{2\tau}-1\right)\D_{t_0}\sum_ih_i^{(1)}e_{i\alpha}^2-
	\frac 1 4\D_{t_0}\sum_i\F_ie_{i\alpha}^2 + 
	\frac{\epsilon}{4}\partial_{t_1}\sum_i\F_ie_{i\alpha}^2
\end{align}
\end{subequations}
Combining the first and the second order of energy equations 
\eqref{th:energy0},\eqref{th:energy1}
by $\partial_t=\partial_{t_0}+\epsilon\partial_{t_1}$, we have,
\begin{align}	\label{th:energy2}
  \partial_t\left(\frac{D}{2}\rho RT +\frac{1}{2}\rho u_\alpha^2\right)=&-
	\frac{1}{2}\partial_\beta\left(\rho u_\alpha^2 u_\beta+
	\frac{(2+D)}\rho RT u_\beta\right) +\nonumber\\
	&\frac1 2\left(\frac{1}{2\tau}-1\right)\epsilon
	\D_t\sum_ih_i^{(1)}e_{i\alpha}^2 +
	\frac{3}{c^2}\rho\varepsilon F_\beta u_\beta
\end{align}
In the equation above, a term with the second-order of $\epsilon$:
$\frac{\epsilon^2}{4}\partial_{t_1}\sum_i\F_ie_{i\alpha}^2$ is neglected.
By using the first order of continuity and momentum equations \eqref{th:continuity0},
\eqref{th:momentum0} we have
\begin{equation}\label{th:ru2}
  \partial_t\left(
\frac 1 2(\rho u_\alpha^2)\right) = -u_\alpha\partial_\alpha P+
	\frac 1 2 u_\alpha^2\partial_\alpha(\rho u_\alpha)-
	u_\alpha\partial_\beta(\rho u_\alpha u_\beta) + \Order(\epsilon)
\end{equation}
where $P$ stands for the full effective pressure given by 
$\partial_{\alpha}P = \partial_{\alpha}(\rho RT)
-\frac{1}{\Upsilon^{(2)}}\rho RT F_{\alpha}$.
\\\\
Subtituting equation \eqref{th:ru2} back into \eqref{th:energy2}, we have,
\footnote{
From here on the calculations have not been rechecked
thoroughly. The reader beware that the
likelihood of accidental is not negligable,
and, in particular, that there may be some
confusion below in the use of
$RT$ and $\varepsilon$.}

\begin{align}	\label{th:energy3}
  \partial_t(\frac{D}{2}\rho RT )+\partial_\beta(\rho\varepsilon u_\beta) = -&
	P\partial_\beta u_\beta+
	\frac1 2\left(\frac{1}{2\tau}-1\right)\epsilon
	\D_t\sum_ih_i^{(1)}e_{i\alpha}^2 +
	\nonumber\\
	&\frac{3}{c^2}\rho\varepsilon F_\beta u_\beta
\end{align}
With the assumtion that, the local energy is conservative. the second term in r.h.s 
of equation above can be rewriten as,
\begin{align}
  \D_t\sum_ih_i^{(1)}e_{i\alpha}^2 = -\tau\partial_\beta
	\left(\partial_t\sum_ih_i^{eq}e_{i\alpha}^2e_{i\beta}+
	\partial_\gamma\sum_ih_i^{eq}e_{i\alpha}^2e_{i\beta}e_{i\gamma}-
	\sum_i\F_ie_{i\alpha}^2e_{i\beta}\right) \nonumber
\end{align}
The high-order terms of $\epsilon$ can be neglected as same as \eqref{eqn:Pih},
\begin{align}
  \D_t\sum_ih_i^{(1)}e_{i\alpha}^2 = -\tau\partial_\beta\left(
	\underbrace{\partial_t\sum_if_i^{eq}e_{i\alpha}^2e_{i\beta}}_{d\Pi_\beta^{(3)}}+
	\underbrace{\partial_\gamma\sum_if_i^{eq}e_{i\alpha}^2e_{i\beta}e_{i\gamma}}
	_{d\Pi_{\beta\gamma}^{(4)}}-
	\sum_i\F_ie_{i\alpha}^2e_{i\beta}\right) \nonumber
\end{align}
Some terms in the equation above can be further simplified as:
\begin{align} \label{th:ru3}
  \partial_t(\rho u_\alpha^2 u_\beta) = &
	u_\alpha u_\beta\partial_t(\rho u_\alpha)+u_\alpha u_\beta[
	\partial_t(\rho u_\alpha)-u_\alpha\partial_t\rho]+
	u_\alpha^2[\partial_t(\rho u_\beta)-u_\beta\partial_t\rho] \nonumber \\
      =&-\partial_\gamma(\rho u_\alpha^2 u_\beta u_\gamma) -
	2u_\alpha u_\beta\partial_\alpha P -
	u_\alpha^2\partial_\beta P + \nonumber \\
	&\frac {12}{c^2D}\rho\varepsilon u_\alpha u_\beta F_\alpha +
	\frac {6}{c^2D}\rho\varepsilon u_\alpha^2 F_\beta
\end{align}
\begin{align} \label{th:reu}
  \partial_t(\rho\varepsilon u_\beta) = &
	u_\beta\partial_t(\rho\varepsilon) +
	\varepsilon[\partial_t(\rho u_\beta)-u_\beta\partial_t\rho] \nonumber \\
  = &	-\partial_\alpha(\rho\varepsilon u_\alpha u_\beta)-
	\frac 2 D \rho\varepsilon u_\beta\partial_\alpha u_\alpha -
	\frac 2 D \varepsilon\partial_\beta(\rho\varepsilon) + \nonumber \\
	&\frac{6}{c^2D}\rho\varepsilon F_\alpha u_\alpha u_\beta  +
	\frac{6}{c^2D}\rho\varepsilon^2 F_\beta
\end{align}
By using the equations \eqref{th:ru3}, \eqref{th:reu} and \eqref{th:Fi-third-moment},
we have,
\begin{align}	\label{th:dPi3a}
  \D_t\sum_ih_i^{(1)}e_{i\alpha}^2 = -\tau\partial_\beta\Big[
	\frac 4 D \rho\varepsilon u_\alpha(\partial_\beta u_\alpha+\partial_\alpha u_\beta) +
	\frac{4(D+2)}{D^2}\rho\varepsilon\partial_\beta\varepsilon - \nonumber \\
	\frac{8}{D^2}\rho\varepsilon u_\beta\partial_\alpha u_\alpha-
	\frac{6(D+4)}{c^2D^2}\rho\varepsilon^2F_\beta+
	\frac{6}{c^2D}\rho\varepsilon F_\alpha u_\alpha u_\beta
	\Big]
\end{align}
Subtituting back equation \eqref{th:dPi3a} into \eqref{th:energy3}, 
we have the enery equation,
\footnote{
Footnote added: the equation of heat transfer in
an incompressible fluid is (compare
\cite{landau}, eq. 50.2)
\begin{equation}
\rho D_t T = (\kappa/c_p)\Delta T
+ (\mu/2c_p)\left(\partial_{\alpha}v_{\beta} + \partial_{\beta}v_{\alpha}
\right)^2
\end{equation}
where $c_p$ is the specific heat at constant pressure.
Equation \eqref{th:dPi3a} definitely contains more
terms than this. Among these are terms of the right
hand side of the energy equation
\eqref{eq:internal-energy-diffuse-interface}, but also
other terms. The meaning and validity of
equation \eqref{th:dPi3a} is unclear. The calculations
of this section however well illustrate that 
constracting thermal multiphase LBE models is not
trivial. 
} 
\begin{align}
  \partial_t(\rho\varepsilon)+\partial_\beta(\rho\varepsilon u_\beta) = -
	P\partial_\beta u_\beta+
	\partial_\beta(\kappa\partial_\beta\varepsilon)+
	\partial_\beta[\mu(\partial_\beta u_\alpha+\partial_\alpha u_\beta)u_\alpha]+ 
	\nonumber \\
	\partial_\beta(\lambda u_\beta\partial_\alpha u_\alpha) + 
	\frac{3}{2}(\mu-\lambda)\partial_\beta(\varepsilon F_\beta)+
	\frac{3}{2}\mu\partial_\beta(F_\alpha u_\alpha u_\beta) +
	\frac{3}{c^2}\rho\varepsilon F_\beta u_\beta
\end{align}
where,
\begin{align}
  \mu = & \frac 2 D \rho\varepsilon\left(\tau-\frac 1 2\right) \\
  \lambda = & -\frac{4}{D^2}\rho\varepsilon\left(\tau-\frac 1 2\right) \\
  \kappa = & \frac{2(2+D)}{D^2}\rho\varepsilon\left(\tau-\frac 1 2\right)
\end{align}
\section{Optimizing a diffuse interface model}
\label{s:optimization}
\begin{figure*}[!hpt]
  \centering
  \includegraphics[height=5cm, width=7cm]{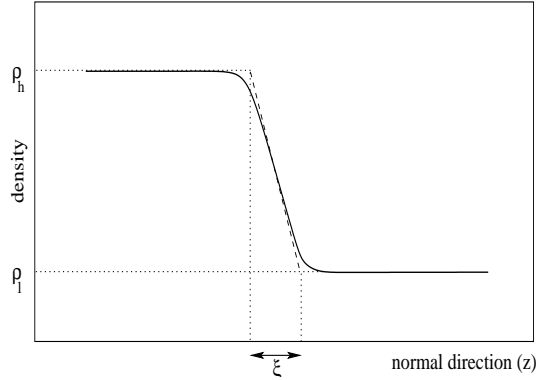}
  \caption{Density profile}
  \label{fig:density.profile}
\end{figure*}

\begin{figure*}[!hpt]
  \centering
  \includegraphics[height=5cm, width=6cm]{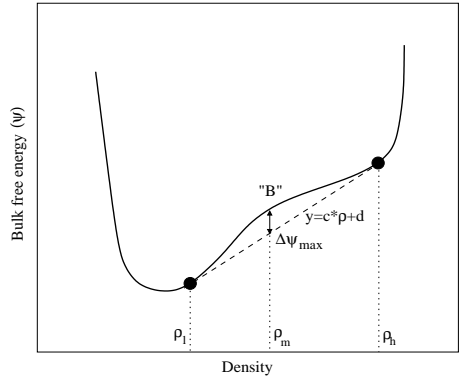}
  \caption{A typical bulk free energy for a one component system}
  \label{fig:psi.original}
\end{figure*}

\begin{figure*}[hpt]
  \centering
  \includegraphics[height=4.2cm, width=5.5cm]{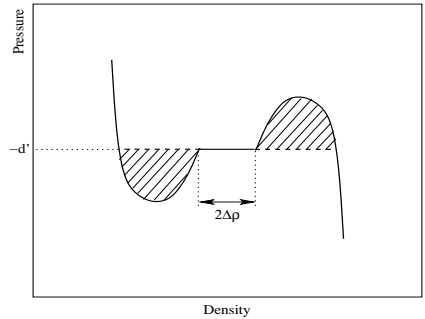}
  \caption{A modification of equation of state}
  \label{fig:press.modified}
\end{figure*}

\begin{figure*}[!hpt]
  \centering
  \includegraphics[height=4.5cm, width=8cm]{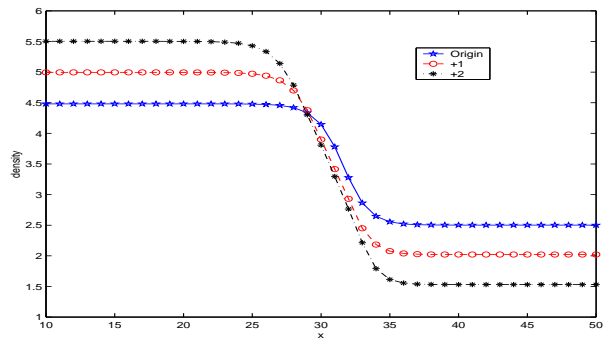}
  \caption{Solutions with the different modification of $\Delta\rho=0,1,2$}
  \label{fig:psi.solution}
\end{figure*}

We have seen that a diffuse interface theory
of a multiphase flow involves a free energy density,
which in the isothermal situation is a function
of density. For a real fluid this energy density is
an observable quantitity in a homogeneous phase.
In the interface, the real form of the 
free energy density is less constrained by experimental
data, and one has some freedom of choice in the model.
\\\\
The numerical difficulty in solving a 
problem of a multiphase flow 
with an interface
by a Lattice Boltzmann, or other diffuse interface method, 
depends on the interface thickness $\xi$, since
the grid size $\Delta x$ cannot be less than $\xi$
\footnote{We assume a homogeneous grid, although
that is not necessary in the Lattice Boltzmann method.
An adaptive grid would involve some sort of front-tracking,
and lead to problems outside the scope of this paper.}.
A typical density profile in an interface is shown
in Fig. \ref{fig:density.profile}.
The interface thickness is approximatively related to the
density gradient and the density difference between
the two phases as
\begin{equation}
	\label{eq:xi-def}
  \frac{d\rho}{dz} \sim \frac{\rho_h-\rho_l}{\xi}
\end{equation}
In a typical application
$(\rho_h-\rho_l)$ is something we would like
to keep fixed. Similarly, surface tension,
and the physical properties of the two phases, are
also things we would like to keep fixed.
\\\\
From \eqref{eq:variation-integrated}
and \eqref{eq:surface-tension2} (or \eqref{eq:surface-tension1})
follow that the surface tension can be written
\begin{equation}
	\label{eq:surface-tension-3}
  \sigma = 2\int_{z_-}^{z_+}\left(\rho f - \rho\mu + P\right) dz
\end{equation}
where $z$ is a coordinate in the normal direction to an interface.
The thermodynamic rule that the pressure and the chemical
potential in two coexisting phases are equal, implies that
the two phases (at densities $\rho_1$ and $\rho_2$ and
bulk free energy densities $\psi_1$ and $\psi_2$) are
related by the  
the double tangent construction of Fig.\ref{fig:psi.original}.
Let us introduce $\tilde\psi$, a linear interpolation 
of $\psi$, and write
\begin{equation}
\psi(\rho) = \tilde\psi(\rho) + \Delta\psi\qquad
\tilde\psi = \psi_1 + \mu(\rho-\rho_1)
\end{equation}
With $P$ the pressure in one of the homogeneous phases
(for instance the $1$ phase) we have in the interface
region
\begin{equation}
\label{eq:delta-psi-def}
\Delta\psi = \psi - \rho\mu + P
\end{equation}
Equations \eqref{eq:delta-psi-def} and
\eqref{eq:surface-tension-3} 
express that
the excess free energy per
unity of surface area of the interface
is the excess free energy density
in the interface region,
integrated over the normal direction to the surface.
\\\\
By using \eqref{eq:xi-def} we have
\begin{equation}
\sigma \sim \overline{\Delta\psi}\,\xi
\end{equation}
where $\overline{\Delta\psi}$ is the mean excess free energy
density in the interface region. Equation
\eqref{eq:variation-integrated} on the other hand means
 \begin{equation}
\overline{\Delta\psi}\sim K \frac{(\Delta\rho)^2}{\xi^2}
\end{equation}
If the parameters to be kept fixed are $\sigma$ and $\Delta\rho$,
the one to be taken large is $\xi$, and the two parameters
to be freely adjusted are 
$K$ and $\overline{\Delta\psi}$ we have 
\begin{equation}
\overline{\Delta\psi} \sim \frac{\sigma}{\xi} \qquad
K \sim \frac{\sigma\xi}{(\Delta\rho)^2}
\end{equation}
To have $\xi$ large, all else constant, one should therefore
take $\overline{\Delta\psi}$ small and $K$
large in the same proportion.
The optimization problem addressed here is that of maximizing
the density difference $\Delta\rho$, while keeping $\sigma$
and $\xi$ constant.   $\overline{\Delta\psi}$ should
then be held constant, while $K$ should be inversely
proportional to the square of $\Delta\rho$.
\\\\
If we use a standard equation of state,
such as the Van der Waals
\begin{equation}
  P = \frac{\rho RT}{1-b}-a\rho^2 \quad
  \psi = \rho RT\ln\left(\frac{\rho}{1-b\rho}\right)-a\rho^2
\end{equation}
the interface is thin, except close to the critical point,
where however also the density difference between the
two phases is small.
Let us now consider the following 
``stretched'' Van der Waals free energy,
where the free energy curve is a straight line
in the interval $[\rho_m - \Delta\rho, \rho_m + \Delta\rho]$, see
Fig.\ref{fig:psi.original}
\begin{figure*}[!hpt]
  \centering
  \includegraphics[height=5cm, width=6cm]{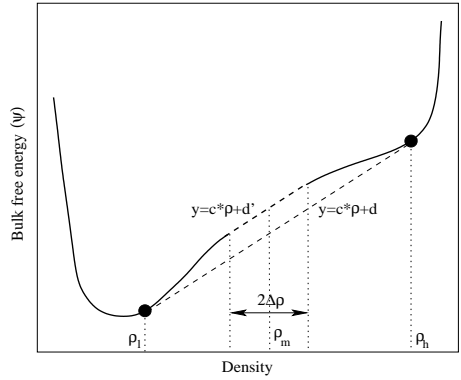}
  \caption{A modification of bulk free energy for a one component system}
  \label{fig_psi.modified}
\end{figure*}
\begin{eqnarray} 
\label{eqex:psi}
\left\{
\begin{array}{ll}
  \psi_{new}(\rho) = \psi(\rho+\Delta\rho) & \rho \le \rho_m-\Delta\rho \\
  \psi_{new}(\rho) = \Delta\psi(\rho_m) +\mu(\rho-\rho_m) & \rho_m-\Delta\rho < \rho < \rho_m+\Delta\rho \\
  \psi_{new}(\rho) = \psi(\rho-\Delta\rho) & \rho \ge \rho_m+\Delta\rho \\
\end{array}
\right.
\end{eqnarray}
Here $\rho_m$ is the point of maximum excess free energy in the
Van der Waals energy, and 
$\Delta\psi(\rho_m)$ is that excess energy. The effect
of the perturbation is hence to move apart the two densities
of the coexisting phases.
\\\\
The pressure in the stretched region is constant, and
equal to the pressure in the two coexisting phases,
see Fig.\ref{fig:press.modified}
With this method, the density difference betweem two
phases can be changed quite easily, see 
 Fig.\ref{fig:psi.solution}
\subsection{Summary}
The form of the free energy in the interface determines
surface tension and interface thickness. In a physical liquid
these two quantities vary together with temperature and pressure
in the phase diagram, and the interface is only thick
close to the critical point. In a numerical procedure that
demands the resolution of the interface, the interface
thickness sets a limit on the smallest grid size that
can be used. 
\\\\
The free energy in the interface can be modified, so as
to e.g. keep surface tension and
interface thickness constant, while increasing
the density difference. From a numerical point of view,
this procedure amounts to smearing the interface,
keeping all else constant, i.e., in some sense
the opposite approach to the use of singular interface
theory.
The advantages of the procedure are that problems 
which are difficult in the singular interface theory
can be readily simulated (see list on page 4), in
particular there is no difficulty with changes in
topology. The drawback is that the simulations will
be of a fictitious liquid with thick interfaces,
and the validity of the simulation results would in
the end have to be established from comparison
with experiments.
\section{List of symbols and conventions used}

\begin{tabbing}
\hspace{2cm} \= \kill
$c_s$ 		\> speed of sound\\
${\bf e}_i$ 	\> lattice (particle) discrete velocities\\
$f_i$ 		\> discrete one-particle distribution function\\
$P,P_{ij}$ 	\> pressure and pressure tensor\\
$R$ 		\> gas constant \\
$T$ 		\> temperature\\
${\bf u}$ 	\> macroscopic velocity\\
$\rho$ 		\> density \\
$\varepsilon_I$	\> internal energy \\
$\varepsilon_K$	\> kinetic energy \\
$\epsilon$	\> expansion parameter \\
$\Order(...)$	\> on the order of ... \\
$w_i$ 		\> weight of different sub-lattices\\
$\alpha,\beta$ 	\> space coordinates\\
$\delta_{\alpha\beta}, \Delta^{(2n)}$
	 	\> Kroenecker delta symbols \\
$\delta t$ 	\> time scale \\
$\delta x$ 	\> length scale \\
$\eta$ 		\> first viscosity \\
$\kappa$ 	\> elastic stress coefficient\\
$\lambda$ 	\> bulk viscosity \\
$\nu$ 		\> kinetic viscosity \\
$\xi$ 		\> second viscosity \\
$\sigma$ 	\> surface tension \\
$\tau$ 		\> relaxation time \\
$\psi$ 		\> bulk free-energy density \\
$\Psi$ 		\> non-local free-energy functional \\
$\Omega_i$ 	\> collision operator \\
${\bf F}$	\> force term \\
$\Pi$		\> momentum flux tensor\\
$\Upsilon^{(2)}$\> Lattice constant, second order tensors \\ 
$\Upsilon^{(4)}$\> Lattice constant, fourth order tensors \\ 
\end{tabbing}
\vspace{1.0cm}
Greek indices ($\alpha,\beta,\ldots$) 
generally label spatial coordinates, while
Latin indices ($i,j,\ldots$) label lattice
vectors.
The Einstein convention of summing repeated spatial
indices has been used throughout the paper. Hence, if
${\bf a} = \{a_1,\ldots,a_D\}$ and ${\bf b} = \{b_1,\ldots,b_D\}$
are two vectors,
then $a_{\alpha}b_{\alpha} = {\bf a}\cdot{\bf b} = \sum_{\alpha=1}^D
a_{\alpha}b_{\alpha}$.
Repeated indices labeling lattice coordinates are
not summed over, unless so indicated.
 
\section*{Acknowledgements}
We thank PSCI for financial support, and Profs.
T. Bohr, A. Nepomnyaschy and S. Succi for comments
and communications. 
\\\\
We thank again Massimo Vergassola for many
valuable discussions, useful advice, and
for taking his time to do the calculations
of sections 4 and 5 above with us.

\newpage

\end{document}